\documentclass[review]{elsarticle}

\usepackage{amsmath,amssymb,amsthm}
\usepackage{multicol}
\usepackage{color,soul}  
\usepackage{lineno,hyperref}
\usepackage{epstopdf}

\usepackage{subfigure}

\usepackage{graphicx}
\usepackage{siunitx}
\usepackage{longtable,tabularx}
\usepackage{subfigmat}
\usepackage{algorithm} 
\usepackage{algpseudocode}

\newcommand\norm[1]{\left\lVert#1\right\rVert} 

\usepackage{xcolor} 

\usepackage{subfigure}

\UseRawInputEncoding

\modulolinenumbers[5]

\journal{Aerospace Science and Technology (AESCTE)}

\begin{document}

\begin{frontmatter}

\title{The Balanced Mode Decomposition Algorithm for\\
        Data-Driven LPV Low-Order Models \\ of Aeroservoelastic Systems}


\author[mymainaddress]{Andrea Iannelli\corref{mycorrespondingauthor}}
\ead{iannelli@control.ee.ethz.ch}

\author[hismainaddress]{Urban Fasel}
\author[mymainaddress]{Roy S. Smith}



\cortext[mycorrespondingauthor]{Corresponding author}
\address[mymainaddress]{Automatic Control Lab, Department of Electrical Engineering, ETH Z\"{u}rich, Switzerland}
\address[hismainaddress]{Department of Mechanical Engineering, University of Washington, Seattle, USA}

\begin{abstract}
A novel approach to reduced-order modeling of high-dimensional systems with time-varying properties is proposed. It combines the problem formulation of the Dynamic Mode Decomposition method with the concept of balanced realization.
It is assumed that the only information available on the system comes from input, state, and output trajectories, thus the approach is fully data-driven. The goal is to obtain an input-output low dimensional linear model which approximates the system across its operating range. Time-varying features of the system are retained by means of a Linear Parameter-Varying representation made of a collection of state-consistent linear time-invariant reduced-order models. The algorithm formulation hinges on the idea of replacing the orthogonal projection onto the Proper Orthogonal Decomposition modes, used in Dynamic Mode Decomposition-based approaches, with a balancing oblique projection constructed from data. As a consequence, the input-output information captured in the lower-dimensional representation is increased compared to other projections onto subspaces of same or lower size. Moreover, a parameter-varying projection is possible while also achieving state-consistency.
The validity of the proposed approach is demonstrated on a morphing wing for airborne wind energy applications by comparing the performance against two recent algorithms. 
Analyses account for both prediction accuracy and closed-loop performance in model predictive control applications.
\end{abstract}


\end{frontmatter}


\section{Introduction}
Data-driven approaches to extract from trajectories of high-dimensional systems, parsimonious models capable of balancing accuracy of the prediction with complexity, are an increasingly popular research topic \cite{Brunton_2020_Review}. In fact, pioneering \emph{ante litteram} contributions to the field, prompted by the goal of identifying low-order structures in complex physical problems such as turbulence, were made in the fluid mechanics and aerodynamics communities \cite{holmes_lumley_berkooz_1996}.
The fundamental idea common to many successful approaches, developed in the wake of these early contributions, is to project the high-dimensional data on a lower dimensional subspace (also constructed from data), such that the most important features of the dynamics are therein preserved. A celebrated example is the Dynamic Mode Decomposition (DMD) approach \cite{schmid_2010,Tu_DMD_2014}, whereby the spectrum of a low-order linear dynamical model approximating the training data is obtained by leveraging the Proper Orthogonal Decomposition (POD) \cite{Berkooz_POD_1993} reduction technique.
Specifically, the projecting subspace provided by POD is spanned by the left singular vectors associated with the largest singular values of a data matrix gathered from observations of the dynamics. The exact interpretation of the largest singular values depend on the inner product used to define the data matrix. In standard applications, where the so-called \emph{snapshot} matrix, corresponding to the correlation matrix between the dynamical states, is used, the largest singular values are associated with the modes capturing most of the energy in the system. Thus, the projection onto the lower dimensional subspace preserves the spatial structures with the highest energy content.
This criterion for choosing the projection subspace might not always give the best results, as low-energy features can have a large effect on the dynamics, e.g. in the case of non-normal systems, which can be found in some fluid dynamics problems \cite{Schmid_book_flows}. Moreover, as recently shown in \cite{Iannelli_ACC21}, projections onto POD modes are not uniquely defined, due to the arbitrariness of the definition of the state. These findings reinforced the known fact that the quality of the approximation highly depends on the choice of inner product and thus care is required when the projection operator is computed.

Despite these potential shortcomings, POD- and DMD-based methods have been successfully applied in various aerospace and control flow problems \cite{Brunton_2020_Review,Noack_interpolation_flow_2007,noack_recursive_2016,LECLAINCHE201588,HU2020106153}. However, a relatively unexplored application domain of data-driven (or equation-free) reduced-order modeling (ROM) is aeroservoelasticity, where the coupling among multiple disciplines (e.g. aerodynamics, structural dynamics) and components of the system (e.g. wing, actuators) often results in high-order models. \begin{color}{black} Expect for a few notable recent exceptions, e.g. the works in \cite{Huang_2015,Huang_2018} where nonlinear ROMs have been developed,\end{color} the standard practice to reduce dimensionality is the use of well established model-based reduction technique \cite{Antoulas_book}. See the work in \cite{Moreno_JA_2014} for an application and further references on this line of research. However, the increasing complexity of the high-fidelity solvers (often made up of distinct sub-solvers for the different disciplines) on one hand, and the potential advantage of recalibrating or directly substituting parts of the code with experimental or flight data on the other, favour the adoption of equation-free strategies.
Among the possible reasons for the lack of their application in the field, two important issues are highlighted here.

First, a common feature of the majority of the available approaches is the focus on internal dynamics, meant here as partial or ordinary differential equations without external excitations and with fully observable states. The work in \cite{Proctor_DMDc} recently extended the DMD framework to controlled systems (DMDc), but the key steps of the algorithm (specifically, the selection of the projecting subspace) do not substantially change. That is, emphasis is not put on preserving the input-output behaviour of the system, which is crucial for control systems.

Second, in aeroservoelastic applications, capturing the variation in the stability and response of the system as the operating conditions change is paramount. This can be done, for example, using the so-called Linear Parameter-Varying (LPV) representation \cite{Toth_LPVbook}, which are of acknowledged benefit for control related tasks \cite{PACKARD199479,BennaniScherer_1998,HE201888,LPVTools}. Unfortunately, obtaining accurate models featuring low orders is notoriously a difficult task \cite{Benner_SIAM_2015}, even for the well explored class of model-based approaches \cite{PousVass_LFTmimo,Ferreres_LPV_TCST,ALJIBOORY201792}. One of the most common strategies is to seek low-order linear time-invariant (LTI) representations for frozen-parameter conditions (defining a grid) and then interpolate them for intermediate values of the parameters.
\begin{color}{black}
Since states, and thus state-space models, are defined up to a nonsingular (similarity) transformation, a correct interpolation requires that the state of each frozen model is defined with respect to the same basis.
\end{color}
The need to work with a consistent state-space basis for the local ROMs, required for a correct interpolation \cite{ZhangLjung_TAC19}, poses a challenge for DMD-inspired data-driven approaches.
State-consistency will depend indeed on the selection of the projecting subspace. If this changes across the parameter range, as it is the case when one computes the POD modes at each grid point, then state-consistency will not hold in general. Conversely, if the subspace is kept fixed for all the frozen-parameter LTI systems, then accuracy might deteriorate since projection will no longer take place onto the optimal (from an energy point of view) subspace for the considered parameter.

Motivated by the discussion above, the main contribution of this paper is the proposal of a novel equation-free approach to obtain LPV low-order models, namely the Balanced Mode Decomposition (BMD) algorithm. The key idea is to use, instead of an orthogonal projection associated with one subspace (as in standard DMD), an oblique projection, which is associated with two subspaces, namely a basis space and a test space, characterizing the range space and null space of the projection, respectively. Oblique projection, often encountered in model reduction \cite{Skelton_Oblique_1987} and system identification \cite{van1996subspace}, was also used for model-based reduction of LPV models in \cite{Theis_TCST2018}.
As detailed in Section \ref{methods:BMD}, the oblique projection proposed here can be interpreted, within the context of DMD-type approaches, as an alternative choice to the subspace spanned by the POD modes, and it is instrumental to achieve two favourable properties.
\begin{color}{black}
The first is that emphasis can be put on the input-output behaviour of the ROM by defining the range and null spaces of the projection as a function of the controllability and observability Gramians. These objects are well known in the context of model order reduction of linear systems, as they are the main ingredients to perform balanced truncation  \cite{Antoulas_book}. This technique consists of transforming the system in balanced coordinates and then removing the states associated with the lowest degree of controllability and observability. 
In the same spirit, range and null spaces of the proposed oblique projection are defined so that
the identified model is (approximately) in balanced coordinates, and thus projection onto lower-dimensional subspaces will preserve the structures in the data matrices that are most observable and controllable. The second favourable property is that the LPV model has a consistent state-space basis \cite{ZhangLjung_TAC19} across the parameter space without having to sacrifice accuracy. This results from the fact that one subspace (the basis space) is common to all parameters and thus provides a common basis. At the same time, the other subspace (the test space) has no influence on the state's basis, and thus can be chosen different for each parameter in order to alleviate the limitations of a fixed subspace projection.
\end{color}

The second contribution of the paper is to extensively compare the results of the BMD method with two recent extensions of DMD with control.
The first algorithm is the algebraic DMDc (aDMDc) \cite{Fonzi_Royal2020}, which extended DMDc to parameter-varying systems described by algebraic, in addition to differential, equations. Including algebraic constraints is very important, for example, when considering state trajectories generated by aerodynamic solvers capturing unsteady effects, such as in panel methods or unsteady vortex lattice methods \cite{MURUA201246}. The second algorithm is the input-output reduced-order model (IOROM) approach, proposed in \cite{Annoni_IJRNC_2017} to construct data-driven reduced-order LPV models. Improved ways of defining the low-dimensional subspace such that state-consistency is achieved while preserving accuracy in the (orthogonal) projection were proposed therein. However, the projection operator is the same for all parameters, and is obtained from the POD modes as in standard DMD. An extension of IOROM to handle algebraic constraints is also developed here in order to allow for a fair comparison. The algorithms are tested on a high-fidelity, fluid-structure interaction (FSI) numerical model of an airborne wind energy (AWE) morphing wing. The FSI simulator is described in \cite{Fasel_AIAA2019} and the wing was analyzed in detail in \cite{Fasel_Scitech2019}. Airborne wind energy and morphing wings are paradigmatic examples of application domains where the system's response originates from complex interactions across different domains, and thus could benefit from equation-free approaches. The first type of comparison investigates the accuracy of the reduced-order models to predict various outputs of the wing as the size of the model is decreased. In a second set of analyses, models featuring different orders are used by a model predictive control (MPC) algorithm to track predefined trajectories of the airborne wind energy system with the goal of gaining insight into the trade-off \begin{color}{black}between size and performance, the latter evaluated by simulating the FSI solver in closed-loop with the MPC controller\end{color}. Preliminary results of the work were presented in \cite{Iannelli_Scitech_2021}.

\begin{figure}[!h]
    \centering
        \includegraphics[width=1\columnwidth]{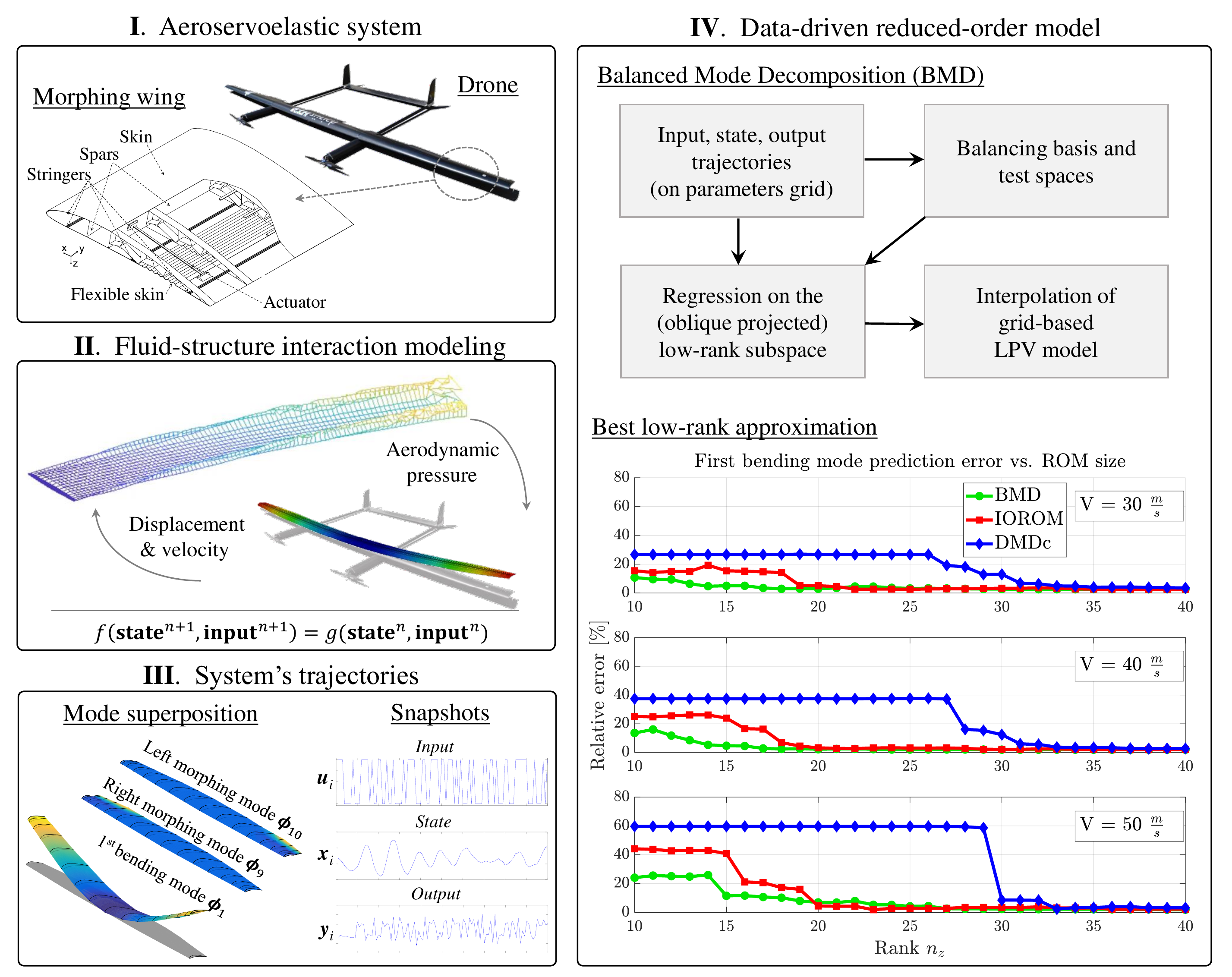}
       \caption{Overview of the algorithm and its proposed application: \textbf{I}. illustrative aeroeservoelastic testcase \cite{Fasel_Scitech2019}; \textbf{II}. typical fluid-structure interaction problem; \textbf{III}. system characterized uniquely by its states, inputs, and outputs trajectories; \textbf{IV}. sketch of the newly proposed BMD algorithm and comparison with two other algorithms.}
        \label{Figure1}
\end{figure}

Figure \ref{Figure1} shows a conceptual representation of the proposed data-driven ROM framework.

\section{Data-driven reduced-order modeling}\label{Algorithms}
This section provides background material on the tools and concepts relevant to the reduced-order modeling algorithm proposed in this work. In Section \ref{methods:state} the general data-driven low-order modeling problem is presented.  Section \ref{methods:aDMDc} reviews the algebraic DMD with Control (aDMDc) \cite{Fonzi_Royal2020}, and Section \ref{methods:IODMD} reports on the input-output reduced-order model (IOROM) \cite{Annoni_IJRNC_2017}. These are the two ROM algorithms from the literature used for comparison in this paper.

\subsection{Problem statement and preliminaries}\label{methods:state}
The starting point is a generic discrete-time nonlinear parameter-varying model which can be used to describe typical control systems, such as aeroservoelastic systems modelled by FSI solvers
\begin{equation}\label{IOROM_in}
    \begin{array}{rcl}
       x_{k+1}  &=&  f(x_k,u_k,\rho_k), \\
       y_k  &=&  h(x_k, u_k, \rho_k),
    \end{array}
\end{equation}
where $x \in \mathbb{R}^{n_x}, u \in \mathbb{R}^{n_u}, y \in \mathbb{R}^{n_y}$ are the state, input and output, and $\rho: \mathbb{N} \rightarrow \mathbb{R}^{n_{\rho}}$ is a vector of time-varying parameters defining the operating conditions of the system. The problem of finding an LPV low-order approximation of (\ref{IOROM_in}) can be divided into two phases: first, \emph{local} LTI approximations for frozen values of $\rho$ in a pre-defined grid $\{\rho^j\}_{j=1}^{n_g}$ are computed; then, an LPV model is obtained through interpolation. The following discussion is concerned with the former phase.

It is assumed that for each frozen value $\bar{\rho}$ there exists an equilibrium (or trim) point characterized by the tuple ($\bar{x}(\bar{\rho})$,$\bar{u}(\bar{\rho})$,$\bar{y}(\bar{\rho})$) such that
\begin{equation*}
    \begin{array}{rcl}
        \bar{x}(\bar{\rho}) &=& f(\bar{x}(\bar{\rho}),\bar{u}(\bar{\rho}),\bar{\rho}),  \\
        \bar{y}(\bar{\rho})  &=& h(\bar{x}(\bar{\rho}),\bar{u}(\bar{\rho}),\bar{\rho}).
    \end{array}
\end{equation*}
The deviation vectors $\tilde{x}_{k} := x_k - \bar{x}(\bar{\rho})$, $\tilde{u}_k := u_k - \bar{u}(\bar{\rho})$, and $\tilde{y}_k := y_k - \bar{y}(\bar{\rho})$ can then be used as states of an LTI approximation of the system around the equilibrium:
    \begin{subequations}\label{IOROM_full}
    \begin{align}
       \tilde{x}_{k+1}  &= A(\bar{\rho})\tilde{x}_{k} + B(\bar{\rho})\tilde{u}_{k},  \label{IOROM_full_1}\\
        \tilde{y}_k &= C(\bar{\rho})\tilde{x}_{k} + D(\bar{\rho})\tilde{u}_{k}, \label{IOROM_full_2}
    \end{align}
\end{subequations}
where ($A(\bar{\rho})$,$B(\bar{\rho})$,$C(\bar{\rho})$,$D(\bar{\rho})$) is a state-space representation completely describing the linearization about the trim point associated with $\bar{\rho}$. 
\begin{color}{black}
The dependence of local (i.e. related to LTI approximations for frozen values of $\rho$) quantities on the parameter $\rho$ will be dropped in the remainder. It is understood that they depend on the particular value of the parameter considered. Instead, the subscript $\rho$ will be used when discussing LPV models.
\end{color}

In the data-driven setting, the only information on the system comes from input, state, and output trajectories $\{x_k,u_{k-1},y_{k-1}\}_{k=1}^{n_s}$ of length $n_s$. After having subtracted from them the corresponding trim values, these trajectories can be used to form the following snapshot matrices
\begin{equation}
\label{equation:IODMD:snapshot_matrices}
    \begin{array}{rcll}
    \vspace{.1in}
       X_0  &=& \begin{bmatrix}x_0 - \bar{x}& x_1 - \bar{x} & ... & x_{n_s-1} - \bar{x}  \end{bmatrix} & \in \mathbb{R}^{n_x \times n_s},  \\ \vspace{.1in}
       X_1  &=& \begin{bmatrix}x_1 - \bar{x}& x_2 - \bar{x} & ... & x_{n_s} - \bar{x}  \end{bmatrix} & \in \mathbb{R}^{n_x \times n_s}, \\ \vspace{.1in}
       U_0  &=& \begin{bmatrix}u_0 - \bar{u}& u_1 - \bar{u} & ... & u_{n_s-1} - \bar{u}  \end{bmatrix}  & \in \mathbb{R}^{n_u \times n_s},\\ \vspace{.1in}
       U_1  &=& \begin{bmatrix}u_1 - \bar{u}& u_2 - \bar{u} & ... & u_{n_s} - \bar{u}  \end{bmatrix}  & \in \mathbb{R}^{n_u \times n_s},\\ \vspace{.1in}
       Y_0  &=& \begin{bmatrix}y_0 - \bar{y}& y_1 - \bar{y} & ... & y_{n_s-1} - \bar{y}  \end{bmatrix} &  \in \mathbb{R}^{n_y \times n_s}. \\ 
    \end{array}
\end{equation}
The notation $\left[ X_0; \;\; U_0\right]$ will denote the operation of stacking row-wise two matrices $X_0$ and $U_0$.

The first goal is to obtain a linear time-invariant low-order approximation of (\ref{IOROM_full}), that is 
\begin{equation*}
    \begin{array}{rcl}
        \tilde{z}_{k+1} &=& F \tilde{z}_k +  G \tilde{u}_k,   \\
        \tilde{y}_k &=& H \tilde{z}_k + D \tilde{u}_k,
    \end{array}
\end{equation*}
where $\tilde{z} \in \mathbb{R}^{n_z}$ 
and $n_z \ll n_x$. Once this is available, the family of frozen LTI systems, or directly the signals of interest, are interpolated so that the response of the system is available at each value $\rho_k$ for a generic time-varying trajectory of the parameter.

\subsection{Algebraic Dynamic Mode Decomposition with Control algorithm}\label{methods:aDMDc}
The algebraic Dynamic Mode Decomposition with Control (aDMDc) algorithm was recently proposed in \cite{Fonzi_Royal2020} to extend the DMDc algorithm to systems described by algebraic-differential equations. The DMDc algorithm from \cite{Proctor_DMDc} is first briefly reviewed. This algorithm seeks a data-driven approximation of the matrices involved in the state equation (\ref{IOROM_full}) by means of two truncated singular value decompositions (SVD) of the snapshot matrices.
The first one is
\begin{equation}\label{Q_DMDc}
   \left[ X_0; \;\; U_0\right] = U \Sigma V^{\top}\cong U_r \Sigma_r V_r^{\top},
\end{equation}
where the subscript $r$ denotes a truncation of the SVD decomposition of order $r$ (obtained by keeping only the $r$ largest singular values in the decomposition). Note that the value of $r$ does not define the size of the final reduced-order model, and it could be set for example by using the hard threshold criterion suggested in \cite{Donoho_thresh_2014}. The effect of choosing $r$ on the accuracy of the model will be discussed in the result section.
The second truncated SVD is computed from the snapshot matrix $X_1$
\begin{equation}\label{Q_DMDc_2}
   X_1 = \hat{U} \hat{\Sigma} \hat{V}^{\top}\cong \hat{U}_{n_z} \hat{\Sigma}_{n_z} \hat{V}_{n_z}^{\top},
\end{equation}
where the columns of $\hat{U}_{n_z}$ are also called POD modes of $X_1$ and are used for the projection onto a lower dimensional space. The selection of $n_z$ defines the size of the reduced-order model. \begin{color}{black}The thresholds used in (\ref{Q_DMDc}) and (\ref{Q_DMDc_2}) should be chosen such that $r>n_z$ \cite{Proctor_DMDc}.\end{color}

An approximation of the high-order matrices appearing in (\ref{IOROM_full}) can be formulated in terms of the truncated SVD (\ref{Q_DMDc}) and the snapshot matrix $X_1$ as
\begin{equation}\label{DMDc_step1}
   \left[ A \;\; B \right] = X_1 V_r^{\top} \Sigma_r^{-1}U_r^{\top}.
\end{equation}
Then, a low-order approximation is obtained by projecting (\ref{DMDc_step1}) onto the set of POD modes by making use of (\ref{Q_DMDc_2})
\begin{equation*}
     \left[ F \;\; G \right] = \left[\hat{U}_{n_z}^{\top}  A \hat{U}_{n_z} \;\;\; \hat{U}_{n_z}^{\top}B \right].
\end{equation*}
Therefore, the low-order model obtained by DMDc (which only includes the state equation) is
\begin{equation*}
        \tilde{z}_{k+1} = F \tilde{z}_k +  G \tilde{u}_k,  \\
\end{equation*}
where $\tilde{z} \in \mathbb{R}^{n_z}$ is the state of the low-order model and the high-order state can be recovered by $\tilde{x} = \hat{U}_{n_z} \tilde{z}$.

The aDMDc algorithm \cite{Fonzi_Royal2020} builds on the DMDc approach and addresses the presence of algebraic constraints in the dynamic equations which might arise when considering unsteady aerodynamics features.
Specifically, the morphing wing analyzed in \cite{Fonzi_Royal2020} is described by an FSI solver that implements a 3D panel method with a free evolving wake inspired by the method in \cite{MURUA201246}. This leads to a dependence of the states' evolution on the inputs at the next time step. Therefore, a slightly different starting point from the general one presented in (\ref{IOROM_in}) has to be considered, namely
\begin{equation}\label{aDMDc_in}
    \begin{array}{rcl}
       g(x_{k+1},u_{k+1})  &=&  f(x_k,u_k,\rho_k), \\
       y_k  &=&  h(x_k, u_k, \rho_k),
    \end{array}
\end{equation}
where $g$ is in general a nonlinear function taking into account the dependence of the states on the control inputs at the next time step. This dependence results from algebraic equations relating the doublet strengths (aerodynamics states) and downwash (function of the other states and the control inputs). This effect is sometimes accounted for with artificial aerodynamic states by simply changing the feedthrough matrix to the outputs. However, to correctly capture the evolution of the states it is important to formulate the problem as stated in (\ref{aDMDc_in}). The reader is referred to \cite{Fonzi_Royal2020} for further discussion on this aspect.

The proposed LTI representation of the system accounting for the algebraic constraints due to the unsteady aerodynamics is
\begin{equation*}
       \tilde{x}_{k+1} = A\tilde{x}_{k} + B\tilde{u}_{k}+R \tilde{u}_{k+1},
\end{equation*}
where, as in DMDc, the objective is to find a low-order approximation for the state equation only. 

The only difference with respect to DMDc is that now the first SVD decomposition is computed with respect to the snapshot matrices $X_0$, $U_0$, and $U_1$, that is
\begin{equation*}
   \left[ X_0; \;\; U_0; \;\; U_1\right] = U \Sigma V^{\top}\cong U_r \Sigma_r V_r^{\top}.
\end{equation*}
And the high-order matrices are thus approximated by
\begin{equation*}
   \left[ A \;\; B \;\; R \right] = X_1 V_r^{\top} \Sigma_r^{-1}U_r^{\top}.
\end{equation*}
A low-order approximation is then obtained by projecting (\ref{DMDc_step1}) onto the same set of POD modes used in DMDc (\ref{Q_DMDc_2})
\begin{equation*}
   \left[ F \;\; G \;\; L \right] =
   \left[\hat{U}_{n_z}^{\top}  A \hat{U}_{n_z} \;\;\; \hat{U}_{n_z}^{\top}B\;\;\; \hat{U}_{n_z}^{\top}R \right].
\end{equation*}
This procedure results in the aDMDc low-order model
\begin{equation}\label{aDMDc_ROM_f}
      \tilde{z}_{k+1} = F\tilde{x}_{k} + G\tilde{u}_{k}+L \tilde{u}_{k+1},
\end{equation}
where the high-order state can again be obtained from $\tilde{x} = \hat{U}_{n_z} \tilde{z}$.

The approach proposed in the parametrically varying version of the aDMDc algorithm is to use a different set of POD modes for each value of $\rho$ in the grid $\{\rho^j\}_{j=1}^{n_g}$. The frozen LTI models (\ref{aDMDc_ROM_f}) are then simulated simultaneously, the relative states are lifted to the high-order ones using the corresponding projection matrices (e.g. $\hat{U}_{n_z}(\rho^j)$ for the model corresponding to the $j$th element in the parameter space), and the state corresponding to the desired value of $\rho$ is obtained by interpolating the high-dimensional states. A first consequence of this approach is that the frozen LTI models  (\ref{aDMDc_ROM_f}) do not have a consistent basis for the state, \begin{color}{black}because the basis of the state is determined by the POD modes, which change across the parameter grid.\end{color} While this has the advantage of projecting over POD modes specifically computed for a particular value of $\rho$, it also requires running in parallel all of the low-order models. Moreover, this algorithm does not provide an LPV model and thus the use of LPV robust control design strategies is precluded \cite{LPVTools}. While other control techniques, such as model predictive control, can still be successfully used (see Section \ref{Results_MPC}), the necessity to run in parallel, multiple low-order models, is a drawback of the method when targeting real-time applications. 

\subsection{Input-output Reduced-Order Model algorithm}\label{methods:IODMD}
The input-output reduced-order model (IOROM) algorithm was proposed in \cite{Annoni_IJRNC_2017} to compute a family of state-consistent data-driven low order LTI state-space models (including the output equation) which can be directly parameterized by the vector $\rho$. 

Consider first the case when there is no parameter dependence (or equivalently, $\rho$ is fixed). Drawing inspiration from the interpretation of DMD as linear dynamics fitting \cite{Tu_DMD_2014}, the main idea is that, given the snapshot matrices (\ref{equation:IODMD:snapshot_matrices}), the matrices ($A$, $B$, $C$, $D$) defining (\ref{IOROM_full}) can be obtained by solving the following least-squares problem
\begin{equation}\label{IOROM_LS_high-nx}
    \min_{A, B, C, D} \norm{\begin{bmatrix}X_1 \\ Y_0 \end{bmatrix} -  \begin{bmatrix}A & B \\ C & D \end{bmatrix}\begin{bmatrix} X_0 \\ U_0 \end{bmatrix}}_F^2,
\end{equation}
where the subscript $F$ denotes the Frobenius norm of a matrix. Without appropriate regularization, this problem would be ill-posed for high-dimensional systems ($n_x \gg 1$). Most importantly, even if (\ref{IOROM_LS_high-nx}) was solved accurately, it would not provide a low dimensional representation of the system. For these reasons, an orthogonal projection of the state onto a low dimensional subspace of dimension $n_z$ is performed by introducing the projection matrix $Q \in \mathbb{R}^{n_x \times n_z}$, where $Q^{\top} Q = I_{n_z}$, such that the orthogonal projection of $\tilde{x}$ on an $n_z$-dimensional subspace is given by $Q Q^{\top} \tilde{x}$. Equivalently, one can think that the original state is approximated by $\tilde{x} \cong Q \tilde{z}$ for some reduced-order state (or coefficient vector) $\tilde{z} \in \mathbb{R}^{n_z}$. This results in the following low-order state-space model
\begin{equation}\label{IOROM_ROM}
    \begin{array}{rcl}
        \tilde{z}_{k+1} &=& (Q^{\top} A Q) \tilde{z}_k + (Q^{\top} B) \tilde{u}_k,   \\ 
        \tilde{y}_k &=& (CQ) \tilde{z}_k + D \tilde{u}_k. 
    \end{array}
\end{equation}
The vector $\tilde{z} = Q^{\top} \tilde{x}\in \mathbb{R}^{n_z}$ can thus be interpreted as the state of the low-rank approximation of (\ref{IOROM_full})
\begin{equation*}
    \begin{bmatrix} A & B \\ C & D\end{bmatrix} \approx \begin{bmatrix} Q F Q^{\top} & Q G \\ H Q^{\top} & D \end{bmatrix} = \begin{bmatrix}Q & 0 \\ 0 & I_{n_y} \end{bmatrix} \begin{bmatrix}F & G \\ H & D \end{bmatrix}\begin{bmatrix}Q^{\top} & 0 \\ 0 & I_{n_u} \end{bmatrix}.
\end{equation*}
The projection matrix $Q$ is constructed from the POD modes of $X_0$, that is
\begin{equation}\label{Q_IOROM}
\begin{aligned}
  Q &= U_{n_z},\\
\textnormal{where}\quad    X_0 &\cong U_{n_z} \Sigma_{n_z} V_{n_z}^{\top}.
    \end{aligned}
\end{equation}

The least-squares problem giving ($F$,$G$,$H$,$D$) is then
\begin{equation}\label{IOROM_LS_prob}
   \min_{F,G,H,D} \norm{\begin{bmatrix}X_1 \\ Y_0 \end{bmatrix} - \begin{bmatrix}Q & 0 \\ 0 & I_{n_y} \end{bmatrix} \begin{bmatrix}F & G \\ H & D \end{bmatrix}\begin{bmatrix}Q^{\top} & 0 \\ 0 & I_{n_u} \end{bmatrix}\begin{bmatrix} X_0 \\ U_0 \end{bmatrix}}_F^2,
\end{equation}
whose solution is
\begin{equation}\label{IOROM_sol}
    \begin{bmatrix} F & G \\ H & D \end{bmatrix}_{opt} = \begin{bmatrix} Q^{\top} X_1 \\ Y_0 \end{bmatrix}\begin{bmatrix} Q^{\top} X_0 \\ U_0 \end{bmatrix}^\dagger,
\end{equation}
where $\dagger$ denotes the pseudo-inverse of a matrix. It is worth noting that the reduced-order model given by the IOROM algorithm is qualitatively similar to the one associated with DMDc.
The main difference (besides the output equation, not considered in DMDc) is that the pseudo-inverse operation, which also amounts to an SVD decomposition and thus is conceptually similar to (\ref{Q_DMDc}), is done here directly on the projected snapshot matrices. This is different than what is done in DMDc, where the SVD decomposition (\ref{Q_DMDc}) is applied to $X_0$ and $U_0$.
A minor difference is also that the POD modes are computed here with respect to $X_0$ instead of $X_1$.

In the parameter-varying case, the regression problem (\ref{IOROM_sol}) is solved at each value of the parameter grid $\{\rho^j\}_{j=1}^{n_g}$ by taking the corresponding snapshot matrices $\{X_0(\rho^j)$, $X_1(\rho^j)$, $U_0(\rho^j)$, $Y_0(\rho^j)\}_{j=1}^{n_g}$. By always using the same projection matrix $Q$ when computing the low-order models at different $\rho$, state-consistency is automatically guaranteed because the orthogonal projection has the same range space. An LPV reduced-order model is then obtained by interpolating (\ref{IOROM_sol}) across the parameter's range. That is
\begin{equation*}
    \begin{array}{rcl}
        \tilde{z}_{k+1} &=& F_{\rho_k} \tilde{z}_k +  G_{\rho_k} \tilde{u}_k + (\bar{z}(\rho_k) - \bar{z}(\rho_{k+1})),  \\
        \tilde{y}_k &=& H_{\rho_k}\tilde{z}_k + D_{\rho_k} \tilde{u}_k,
    \end{array}
\end{equation*}
where ($F_{\rho_k}$,$G_{\rho_k}$,$H_{\rho_k}$,$D_{\rho_k}$) are obtained by interpolating the corresponding matrices for the value of $\rho$ at timestep $k$. Note that the low-order state is $\tilde{z}_{k} = z_k - \bar{z}(\rho_k)$, where the trim point $\bar{z}(\rho_k) = Q^{\top} \bar{x}(\rho_k)$ can change as a function of $\rho$. The term $(\bar{z}(\rho_k) - \bar{z}(\rho_{k+1}))$ is added to correctly take into account this effect \cite{Annoni_IJRNC_2017}.

Since the choice of a fixed projection matrix is typically associated with less accuracy, two strategies are proposed in \cite{Annoni_IJRNC_2017} to alleviate this issue. The first consists of using in the decomposition (\ref{Q_IOROM}) a \emph{fat} snapshot matrix $X_0$ obtained by stacking column-wise the snapshot matrices of multiple parameters. This matrix will feature $n_s n_g$ columns, which can result in computationally expensive calculations when this number is large. The second, less accurate but more practical in case several grid points are analyzed, consists of iteratively building $Q$ by incrementally processing the snapshot data from each grid point in a similar fashion to the Gram-Schmidt orthogonalization procedure. The former strategy is used here when showing results for the IOROM algorithm, together with a linear interpolation of the state-space matrices.

\section{Balanced Mode Decomposition with oblique projection algorithm}\label{methods:BMD}
This section presents the technical aspects of the Balanced Mode Decomposition (BMD) algorithm proposed in this paper.
Section \ref{BMD:connections} clarifies the goals and the novelty of the contribution with respect to previous works. Section \ref{BMD:Algorithm} presents the algorithm and Section \ref{BMD:basis} details the computation of the subspaces defining the oblique projection. Finally, Section \ref{BMD:algeb} presents a version of the algorithm which can handle algebraic constraints and thus allows the analyses in Section \ref{Results} of the morphing wing with an unsteady aerodynamics model.

\subsection{Novelty and connections with prior work}\label{BMD:connections}
The main motivation for the proposal of the BMD algorithm for data-driven LPV low-order modeling is to address two limitations of recent extensions of the celebrated DMD method to input-output parameter-varying models. The first one concerns the use of $Q$ (i.e. the subspace spanned by the most energetic POD modes according to the standard choice of inner product as unweighted scalar product) for the projection of the higher-order dynamics, which is suboptimal as also acknowledged by the authors of \cite{Annoni_IJRNC_2017}.
In the input-output context, a subspace typically providing lower input-output errors with respect to the others having same size $n_z$ is the one where the system's state is in balanced coordinates \cite{Moore_Bal_1981}. This is indeed the rationale behind balanced truncation, which consists of removing the states corresponding to the smallest $n_x-n_z$ Hankel singular values \cite{Glover_IJC_1984}. The justification for this is that the sum of the Hankel singular values provides a lower bound, and for systems in balanced coordinates, an upper bound on the error of the approximation achieved by removing system's states. Even though not guaranteed to be optimal, balanced truncation is a very effective tool in model-based order reduction \cite{Antoulas_book,Himpe_gramians}. These ideas are used here to propose a new projection operator for the high-dimensional state.



Whereas the aspect mentioned above is relevant also for the case where the sought model is an LTI (i.e. when there is no parameter dependence), the second one is specific to the LPV setting. Precisely, the second limitation addressed by the BMD algorithm concerns how to handle state-consistency across the frozen models in order to estimate the system's response at intermediate points in the parameter grid. In the currently available approaches, this is addressed in two possible ways. When state-consistency is not fulfilled, all reduced-order models are run in parallel by interpolating directly the high-dimensional lifted state. This is the case of aDMDc, and while it has the advantage that the projection operators are parameter-dependent (i.e. at each parameter's value one can use a different set of POD modes), an LPV model is not available and moreover computational efficiency might be compromised. On the other hand, a parameter-independent projection matrix for all frozen models can be used in order to guarantee state-consistency. This is the case for the IOROM algorithm, and it has the drawback that, whereas the orthonormal basis associated with the $n_z$ most energetic modes will be in general different at each value of $\rho$, a fixed one that is common to all parameters is used.

The central idea to overcome both of the aforementioned issues is to replace the orthogonal projection employed in standard POD-based approaches by an oblique projection. Given $V \in \mathbb{R}^{n_x \times  n_z}$ and $W \in \mathbb{R}^{n_x \times n_z}$, such that $W$ is bi-orthogonal to $V$, i.e. $W^{\top} V = I_{n_z}$, the oblique projection of $\tilde{x}$ is given by $\Pi \tilde{x}$, where $\Pi=V W^{\top}$.
As a result, the original state is approximated as $\tilde{x} \cong V \tilde{z}$ (where, as before, $\tilde{z} \in \mathbb{R}^{n_z}$ is the reduced-order state), and the component of $\tilde{x}$ that is eliminated by the projection is in the nullspace of $\Pi$. As opposed to the orthogonal projection, which is characterized by a single subspace (the one spanned by the columns of $Q$), the oblique projection is defined by two subspaces: the basis space (spanned by the columns of $V$), such that the projection of $\tilde{x}$ lies in the span of $V$; and the test space (spanned by the columns of $W$), such that the projection $V \tilde{z}$ has zero error within it, i.e. $W^{\top}\left(\tilde{x}-V \tilde{z} \right)=0$. Technically, the high-dimensional state vector is projected along the orthogonal complement of the subspace spanned by the columns of $W$ onto a subspace spanned by the columns of $V$. In practice, this means that what is lost by projecting $\tilde{x}$ (i.e. the nullspace of the projection) is orthogonal to $W$, and the state basis only depends on $V$.
The two issues discussed above are then addressed by: computing $V$ and $W$ from the \emph{empirical} controllability and observability Gramians of the system (which leads to a model-free balanced truncation); employing a fixed $V$ and a parameter-dependent $W$. \begin{color}{black} Since $V$ by definition defines the basis of the vector space where the state of each model is defined, this basis will be common to all the local state-space models, and thus state-consistency is guaranteed.\end{color}

The idea of using an oblique projection for LPV model-order reduction was first proposed in \cite{Theis_TCST2018}. Therein, the setting where a model of the system is available (in the form of high-order state-space models) is considered, and thus both the construction of $V$ and $W$, and the computation of the low-order model, is model-based. In the data-driven ROM literature, balancing concepts are used in two important techniques, namely Balanced POD (BPOD) \cite{Rowley_BPOD_2005} and the Eigensystem Realization Algorithm (ERA) \cite{Ma_ERA_2011}.
The former is only partially equation-free: the controllability Gramian is computed from data, while for the observability Gramian an adjoint simulation model is needed. Additionally, the high-order state-space matrices are required for the balanced projection.
For the case of ERA, a balanced model comes from impulse response simulations of the model in the spirit of system identification algorithms from realization theory \cite{Viberg_Auto1995}. The ERA algorithm is closely related to BPOD, as it can be interpreted as a data-driven balanced truncation. An important difference is that ERA provides only the reduced-order model and not the balancing transformation, namely the set of vectors known as balancing and adjoint modes in BPOD. These modes are the counterpart of the basis and test space in BMD, respectively, and are a desirable output of a ROM algorithm as they show the most important spatio-temporal structures in the dynamics. In an aeroservoelastic setting, this can provide insights into efficient design solutions.
\begin{color}{black}
It is recalled that BPOD can be interpreted as a special case of POD when impulse responses are used to build the snaphost matrices and the observability Gramian is used as inner product \cite{Rowley_BPOD_2005}. Conceptually (because in practice the algorithm formulation is articulated in a different way), it can be helpful to think of BMD as a version of DMD that makes use of this special case of POD.
\end{color}

\subsection{BMD regression problem}\label{BMD:Algorithm}
We consider first the frozen-parameter case and, by virtue of the previously discussed oblique projection, propose the following low-order LTI system model
\begin{equation}\label{BMD_ROM_eq}
    \begin{array}{rcl}
        \tilde{z}_{k+1} &=& (W^{\top} A V) \tilde{z}_k + (W^{\top} B) \tilde{u}_k,   \\ 
        \tilde{y}_k &=& (CV) \tilde{z}_k + D \tilde{u}_k, 
    \end{array}
\end{equation}
where the computation of the balancing basis $V$ and test spaces $W$ from system's trajectories will be detailed in Section \ref{BMD:basis}. The vector $\tilde{z} = W^{\top} \tilde{x}\in \mathbb{R}^{n_z}$ can thus be interpreted as the state associated with the following low-rank approximation of (\ref{IOROM_full})
\begin{equation}\label{BMD_rankApp}
    \begin{bmatrix} A & B \\ C & D\end{bmatrix} \approx \begin{bmatrix} V F W^{\top} & V G \\ H W^{\top} & D \end{bmatrix} = \begin{bmatrix}V & 0 \\ 0 & I_{n_y} \end{bmatrix} \begin{bmatrix}F & G \\ H & D \end{bmatrix}\begin{bmatrix}W^{\top} & 0 \\ 0 & I_{n_u} \end{bmatrix}.
\end{equation}
The matrices ($F$,$G$,$H$,$D$) can then be obtained with the following least-squares problem
\begin{equation}\label{BMD_LS}
   \min_{F,G,H,D} \norm{\begin{bmatrix}X_1 \\ Y_0 \end{bmatrix} - \begin{bmatrix}V & 0 \\ 0 & I_{n_y} \end{bmatrix} \begin{bmatrix}F & G \\ H & D \end{bmatrix}\begin{bmatrix}W^{\top} & 0 \\ 0 & I_{n_u} \end{bmatrix}\begin{bmatrix} X_0 \\ U_0 \end{bmatrix}}_F^2,
\end{equation}
which has solution
\begin{equation}\label{BMD_sol}
    \begin{bmatrix} F & G \\ H & D \end{bmatrix}_{opt} = \begin{bmatrix} W^{\top} X_1 \\ Y_0 \end{bmatrix}\begin{bmatrix} W^{\top} X_0 \\ U_0 \end{bmatrix}^\dagger.
\end{equation}

To build a low-order LPV model, snapshot matrices are first collected for the values of the parameter in the grid $\{\rho^j\}_{j=1}^{n_g}$, and the least-squares problem (\ref{BMD_LS}) is solved at each grid point. Crucially, the test space $W$ is allowed to be a function of $\rho$. This leads to the following solution for the reduced-order models in the grid
\begin{equation}\label{BMD_LPV_sol}
\begin{bmatrix} F(\rho^j) & G(\rho^j) \\ H(\rho^j) & D(\rho^j) \end{bmatrix}_{opt} = \begin{bmatrix} W^{\top}(\rho^j)X_1(\rho^j) \\ Y_0(\rho^j) \end{bmatrix}\begin{bmatrix} W^{\top}(\rho^j)X_0(\rho^j) \\ U_0(\rho^j) \end{bmatrix}^\dagger,
\end{equation}
where the dependence on $\rho$ of the local quantities has been here explicitly reported in order to clearly point out that all the objects involved in the construction of the low-order state-space models are parameter-varying.

The BMD LPV reduced-order model is then obtained by interpolating the frozen matrices (\ref{BMD_LPV_sol}) across the parameter's range
\begin{equation}\label{BMD_LPV_ROM}
    \begin{array}{rcl}
        \tilde{z}_{k+1} &=& F_{\rho_k} \tilde{z}_k +  G_{\rho_k} \tilde{u}_k + (\bar{z}(\rho_k) - \bar{z}(\rho_{k+1})),  \\
        \tilde{y}_k &=& H_{\rho_k}\tilde{z}_k + D_{\rho_k} \tilde{u}_k,
    \end{array}
\end{equation}
where, as in IOROM, ($F_{\rho_k}$,$G_{\rho_k}$,$H_{\rho_k}$,$D_{\rho_k}$) are obtained by interpolating the corresponding matrices for the value of $\rho$ at timestep $k$, and the term $(\bar{z}(\rho_k) - \bar{z}(\rho_{k+1}))$ takes into account the fact that the equilibrium point associated with each $\rho$ is in general different, and $\bar{z}(\rho_k) = W^{\top} \bar{x}(\rho_k)$. Note that, since $V$ is fixed, the basis space is common to all the frozen models and thus the interpolation can be done at the state-matrices level (as in IOROM). However, the projection is parameter-dependent due to the use of a parameter-varying test space $W(\rho)$.

\subsection{Basis and test spaces construction}\label{BMD:basis}

In order to preserve the most important features of the input-output mapping of the system when projecting into the lower order subspace of dimension $n_z$, the matrices $V$ and $W$ are computed from the controllability and observability Gramians, respectively $W_c$ and $W_o$. This ensures that the projection preserves the most observable and controllable states, enabling an approximate data-driven balanced truncation of the reduced-order LPV model.

Because the approach is fully data-driven, empirical Gramians are computed from data matrices consisting of appropriate state trajectories using known systems theoretical results. The empirical controllability Gramian can be obtained, following the definition \cite{Moore_Bal_1981,Laub_TAC87}, by impulse response simulations (one for each input channel). As for the empirical observability Gramian, if the model is linear and its adjoint is available, then it can be computed from impulse response simulations (one for each output channel) of the adjoint system, as done in balanced POD \cite{Rowley_BPOD_2005}. This computation is identical to the one giving the controllability Gramian, but is applied to the adjoint system (for an LTI model in state-space form, this is obtained by replacing $A$ and $B$ by $A^{\top}$ and $C^{\top}$).
If the above does not hold, for example in case one has only access to the system's trajectories and not to the model's matrices, the approach developed in \cite{Lall_IJRNC_2002}, valid also for nonlinear systems and not requiring an adjoint model, can be used. In this method the data matrices used for the Gramian computation consist of state trajectories
obtained from unforced (zero input) simulations (one for each state) obtained by perturbing the initial condition of each state.
Since these are unforced responses, when the system is sufficiently damped, it will be generally sufficient to observe only the initial time-steps and thus this calculation can be parallelized and efficiently implemented to reduce the computational time.

Once $W_c$ and $W_o$ are available, a procedure based on \cite[Proposition~2]{Theis_TCST2018} is employed to compute the test and basis spaces.
This construction is reported in the first part of the pseudocode given in Algorithm 1, which summarizes input, output, and main steps of the BMD algorithm (MATLAB notation is used for matrix operations and rows/columns selection). For a fixed value of $\rho$, the construction proposed in \cite{Theis_TCST2018} is an equivalent procedure, but more numerically stable for large-scale systems, to the well known square root algorithm for balanced truncation \cite{Laub_TAC87}. Indeed, it can be noted (see lines 5-11) that the subspace $V$ is taken as a basis for span($L_c \tilde{U}$), where $L_c$ ($L_o$) is a Cholesky factor of $W_c$ ($W_o$) and $\tilde{U}$ consists of the first $n_z$ left singular vectors of $H=L_c^{\top} L_o$. The singular values of $H$ are the Hankel singular values of the system and the SVD decomposition of $H$ plays a fundamental role in balanced reduction \cite{Moore_Bal_1981,Laub_TAC87}.
As for the subspace $W$, it can be shown that the expression in line 16 is equivalent to $W=W_o V(V^{\top}W_o V)^{-1}$, but it is computed with improved numerical robustness \cite{golub2013matrix} by making use of a thin QR factorization (line 13). These choices of $V$ and $W$ are shown \cite[Proposition~2]{Theis_TCST2018} to provide the same balancing projection operator used in the square root algorithm.
\begin{color}{black}
The desired order $n_z$ is indicated as an input of Algorithm 1, because one might want to obtain a low-order model with a specified size. This is also the approach used in the analyses presented in Section \ref{Results}. Alternatively, $n_z$ can be indirectly defined as in aDMDc and IOROM by setting a threshold on the singular values of a data matrix. Whereas in those two methods this is done with respect to singular values of the snapshot matrix of the state ($X_1$ in Eq. (\ref{Q_DMDc_2}) and $X_0$ in Eq. (\ref{Q_IOROM})), in the BMD algorithm the matrix $H$ should be considered. That is, a threshold on the Hankel singular values of the system can be selected, as in standard balanced reduction. This threshold can be conveniently expressed as percentage of the largest Hankel singular value. Note that for each parameter in the grid there is a different matrix $H$ (line 4), and thus, given a threshold, the number of truncated singular values might differ at each grid point. Because obviously a common value of $n_z$ for all the local models should be selected, a remedy for this is e.g. to choose as $n_z$ the maximum number of truncated singular values across the grid.
\end{color}

\begin{algorithm}[!h]
    \caption{Balanced Mode Decomposition with oblique projection}
     \textbf{Input}: parameter grid points $\{\rho^j\}_{j=1}^{n_g}$; snapshot matrices
          $\{X_0(\rho^j)$, $X_1(\rho^j)$, $U_0(\rho^j)$, $Y_0(\rho^j)\}_{j=1}^{n_g}$; empirical Gramians $\{W_c(\rho^j)$, $W_o(\rho^j)\}_{j=1}^{n_g}$;
     desired model order $n_z$. \\ 
\textbf{Output}: test space projection matrices $\{W(\rho^j)\}_{j=1}^{n_g}$;
  fixed basis space projection matrix $V$; reduced-order models at the grid points
$\{F(\rho^j)$,$G(\rho^j)$,$H(\rho^j)$,$D(\rho^j)\}_{j=1}^{n_g}$.

    \begin{algorithmic}[1]
        \For{$j=1,...,n_g$}
     \State $L_c(\rho^j)L_c(\rho^j)^{\top} = W_c(\rho^j)\quad \quad$  \textnormal{Cholesky factorization of }$W_c$
            \State $L_o(\rho^j)L_o(\rho^j)^{\top} = W_o(\rho^j)\quad \quad$ \textnormal{Cholesky factorization of }$W_o$
             \State $H(\rho^j) = L_c(\rho^j)^{\top} L_o(\rho^j)$
            \State $(U,\star,\star) = \textnormal{svd}(H(\rho^j))$
             \State $\tilde{U} = U(:,1:n_z)$
            \State $(\bar{U},\star,\star) = \textnormal{svd}(L_c(\rho^j)\tilde{U})$
            \State $\bar{Q}(:,1 + n_z(j-1):n_z j) = \bar{U}(:,1:n_z)$
        \EndFor

        \State $(U_V,\star,\star) = \textnormal{svd}(\bar{Q})$
        \State $V = U_V(:,1:n_z)\quad \quad \quad \quad\quad \quad $  \textnormal{ Fixed basis space }

        \For{$j = 1,...,n_g$}
            \State $(\tilde{Q},\tilde{R}) = \textnormal{qr}(L_o(\rho^j)^{\top} V)\quad \quad$  \textnormal{Thin QR factorization}
            \State $Q = \tilde{Q}(:,1:n_z)$
          \State $R = \tilde{R}(1:n_z,:)$
           \State $W(\rho^j) = L_o(\rho^j) Q (R^{\top})^{-1}\quad \quad$  \textnormal{Parameter-varying test space}
             \State
             \begin{equation*}
             \begin{bmatrix} F(\rho^j) & G(\rho^j) \\ H(\rho^j) & D(\rho^j) \end{bmatrix} = \begin{bmatrix} W^{\top}(\rho^j)X_1(\rho^j) \\ Y_0(\rho^j) \end{bmatrix}\begin{bmatrix} W^{\top}(\rho^j)X_0(\rho^j) \\ U_0(\rho^j) \end{bmatrix}^\dagger \quad \quad \textnormal{solution of the BMD regression problem}
             \end{equation*}
        \EndFor
    \end{algorithmic}
\end{algorithm}

\newpage

The output $\{F(\rho^j)$,$G(\rho^j)$,$H(\rho^j)$,$D(\rho^j)\}_{j=1}^{n_g}$ provided by the BMD algorithm is a grid LPV model. After an interpolation algorithm to evaluate the matrices' entries for any value of $\rho$ inside the considered range has been chosen, this model can be used for simulation and control design. Note also that recently proposed robust analysis methods for linear-time varying (LTV) systems \cite{Seiler_LTV_FH_UB,Iannelli_JGCD_LTV} can be applied to this model, e.g. to investigate specific aircraft manoeuvres. Indeed, by fixing a particular trajectory for $\rho$ the LPV system is transformed into an LTV one.
Moreover, the parameter-varying test space $W(\rho^j)$ can be useful to gain insights into the aeroservoelastic modes which have been eliminated and those that have been kept in the projection, while the parameter-independent basis space can be used to recover at each time-step $k$ the high-dimensional state via the transformation $\tilde{x}_k = V \tilde{z}_k$.


As noted in the introduction, the algorithm provides an approximate balanced truncation. Approximation is related to the use of empirical Gramians, which are only finite-time approximations of the true ones (for this reason, also called finite-time Gramians) since their computation is trajectory-based. As a result, they only provide in principle a finite-time balanced realization \cite{Antoulas_book}, whereas the theoretical order reduction error bounds are only available for infinite-time balanced realizations. This source of error can however be made arbitrarily small by using long enough data sequences for constructing the Gramians. The slowest decay rate of the system's impulse responses is the key parameter to consider when choosing the length of the trajectory \cite{Markovsky_2005}.


\subsection{Extension to handle algebraic constraints}\label{BMD:algeb}
Since the BMD algorithm will be applied in Section \ref{Results} to a wing described by an FSI solver which implements the algebraic constraints described in Section \ref{methods:aDMDc}, an extension to handle this instance is presented here. 
For a fixed value of $\rho$, the model structure for the high-order model becomes
\begin{equation*}
    \begin{array}{rcl}
         \tilde{x}_{k+1} &=& A \tilde{x}_k + B \tilde{u}_k + R\tilde{u}_{k+1}, \\
         \tilde{y}_k &=& C \tilde{x}_k + D \tilde{u}_k + P \tilde{u}_{k+1},
    \end{array}
\end{equation*}
where a potential effect of the algebraic constraints in the output equation is also considered via the matrix $P$ (in the analyses of the morphing wing this matrix was, as expected, always found to be zero). Therefore, the low-order approximation (for each value of $\rho$ in the parameter grid) becomes
\begin{equation*}
\begin{split}
    \begin{bmatrix} A & B & R \\ C & D & P \end{bmatrix} \approx
    \begin{bmatrix} V F W^{\top}& V G & V L \\ H W^{\top} & D & P \end{bmatrix} = 
    \begin{bmatrix}V & 0 \\ 0 & I_{n_y} \end{bmatrix} \begin{bmatrix}F & G & L \\ H & D & P \end{bmatrix}\begin{bmatrix}W^{\top} & 0 & 0 \\ 0 & I_{n_u} & 0 \\ 0 & 0 & I_{n_y} \end{bmatrix}.
\end{split}
\end{equation*}
The new objective function to be minimized is
\begin{equation*}
\begin{split}
    \min_{F, G, L, H, D,P} \norm{\begin{bmatrix}X_1 \\ Y_0 \end{bmatrix} - \begin{bmatrix}W & 0 \\ 0 & I_{n_y} \end{bmatrix} \begin{bmatrix}F & G & L \\ H & D & P \end{bmatrix}\begin{bmatrix}W^{\top} & 0 & 0\\ 0 & I_{n_u} & 0 \\ 0 & 0 & I_{n_y} \end{bmatrix}\begin{bmatrix} X_0 \\ U_0 \\ U_1 \end{bmatrix}}_F^2,
\end{split}
\end{equation*}
and the new optimal solution is
\begin{equation*}
 \begin{bmatrix}F & G & L \\ H & D & P \end{bmatrix}=\begin{bmatrix}W^{\top} X_1 \\ Y_0 \end{bmatrix}\begin{bmatrix} W^{\top} X_0 \\ U_0 \\ U_1 \end{bmatrix}^\dagger.
\end{equation*}
\begin{color}{black}
The BMD LPV reduced-order model with algebraic constraints is then, in analogy to (\ref{BMD_LPV_ROM}), given by 
\begin{equation}\label{BMD_LPV_ROM_algeb}
    \begin{array}{rcl}
        \tilde{z}_{k+1} &=& F_{\rho_k} \tilde{z}_k +  G_{\rho_k} \tilde{u}_k+  L_{\rho_k} \tilde{u}_{k+1} + (\bar{z}(\rho_k) - \bar{z}(\rho_{k+1})),  \\
        \tilde{y}_k &=& H_{\rho_k}\tilde{z}_k + D_{\rho_k} \tilde{u}_k+  P_{\rho_k} \tilde{u}_{k+1}.
    \end{array}
\end{equation}

The IOROM algorithm has also been extended to the algebraic-differential case in order to allow its application to the test case considered in Section \ref{Results}. This can be done in a similar way to what has been shown above for BMD. Specifically, starting from (\ref{IOROM_LS_prob}), the new least-squares problem for the IOROM reduced-order model is
\begin{equation*}
\begin{split}
    \min_{F, G, L, H, D,P} \norm{\begin{bmatrix}X_1 \\ Y_0 \end{bmatrix} - \begin{bmatrix}Q & 0 \\ 0 & I_{n_y} \end{bmatrix} \begin{bmatrix}F & G & L \\ H & D & P \end{bmatrix}\begin{bmatrix}Q^{\top} & 0 & 0\\ 0 & I_{n_u} & 0 \\ 0 & 0 & I_{n_y} \end{bmatrix}\begin{bmatrix} X_0 \\ U_0 \\ U_1 \end{bmatrix}}_F^2,
\end{split}
\end{equation*}
which has the following solution
\begin{equation}\label{IOROM_LPV_ROM_algeb}
 \begin{bmatrix}F & G & L \\ H & D & P \end{bmatrix}=\begin{bmatrix}Q^{\top} X_1 \\ Y_0 \end{bmatrix}\begin{bmatrix} Q^{\top} X_0 \\ U_0 \\ U_1 \end{bmatrix}^\dagger.
\end{equation}
Eq. (\ref{BMD_LPV_ROM_algeb}), when the interpolated matrices are taken from (\ref{IOROM_LPV_ROM_algeb}), provides the IOROM LPV reduced-order model with algebraic constraints.
\end{color}

\section{Results}\label{Results}

This section presents and discusses results obtained by applying the three algorithms aDMDc, IOROM, and BMD to the flexible and highly cambered morphing wing depicted in Fig. \ref{Figure1}. The wing is made of composite material, and the trailing edges are able to morph and, by doing so, to increase or decrease the camber, thus replacing conventional ailerons. The reader is referred to the related previous works for details on the wing design \cite{Keidel_ICAST2017} and its investigation with fluid-structure interaction tools \cite{Fasel_Scitech2019}.

\subsection{Summary of the wing's FSI model}

The high-fidelity FSI model of the morphing wing is presented in detail in \cite{Fasel_AIAA2019,Fasel_PhD}. The structural model is based on a combination of linear plate and beam elements. The external skin of the lifting surface and the Voronoi-based internal structure are modelled using plates, while the stringers are modelled with beam elements. The stiffness and mass matrices are obtained with the commercial software Nastran \cite{NASTRAN_aeroelastic}. \begin{color}{black}From these, a modal decomposition is performed to extract the structural modes of the wing, which are coupled via thin plate spline and inverse distance weighting \cite{Shepard_IDW_68} with the aerodynamic model.\end{color} The aerodynamic model is based on a 3D unsteady panel method \cite{Katz2001}. The unsteadiness of the flow is represented by shedding at each time step a new row of vortex ring singularities after the trailing edge. All the other wake nodes are then moved, via a second-order Runge–Kutta integration scheme, using the local velocity of the flow. The aerodynamic forces on the surface are computed with the coefficient of pressure on each panel, considering the far field velocity, the induced velocity by the wing itself, and the induced velocity by all the wake panels.

The state of the system $x$ consists of the total number of structural modes of the wing and the doublet strengths (from the 3D panel method solver), with $n_x$=618. The input vector $u$ of size $n_u$=6 is given by
\begin{equation}\label{input_channels}
 u=\left[\alpha; \; \; p; \; \; q; \; \; r; \; \; F_s; \; \; F_{as} \right],
\end{equation}
where $\alpha$ is the angle of attack, $p$, $q$, and $r$ are the roll, pitch, and yaw rotation rates, and $F_s$ and $F_{as}$ are the (normalized) symmetric and anti-symmetric morphing actuation inputs.
Their value is associated with a camber deformation and is thus related to a trailing edge deflection: specifically, the amount of upwards (negative value) or downwards (positive value) deflection.

\begin{color}{black}As for the output channels, we will consider both single output and multiple output models.\end{color} Emphasis will be given to the case where the output is the first bending mode of the wing ($n_y$=1), since this is usually the one associated with dynamic instabilities and large deformations, and thus it is of particular interest for active control tasks \cite{Waitman_JGDC2020,Theis_JGCD2020}.
The flight speed will be considered as the time-varying parameter ($n_{\rho}$=1).

The training phase, common to both the algorithms and consisting of generating the snapshot matrices in Eq. (\ref{equation:IODMD:snapshot_matrices}), is carried out
by exciting the system with a series of impulses deployed in random order in all input channels, following the same procedure adopted in \cite{Fonzi_Royal2020}.
\begin{color}{black} The amplitude of these signals has been chosen so that nonlinear effects due to the wake's evolution are not excited. The same precaution is used for the computation of the empirical gramians.\end{color}
Trajectories are of length $n_s$=500 and are recorded with sampling time 0.006 s.


\subsection{Fixed-parameter models}\label{Results_fixedV}
\begin{figure}[!h]
    \centering
        \includegraphics[width=1\columnwidth]{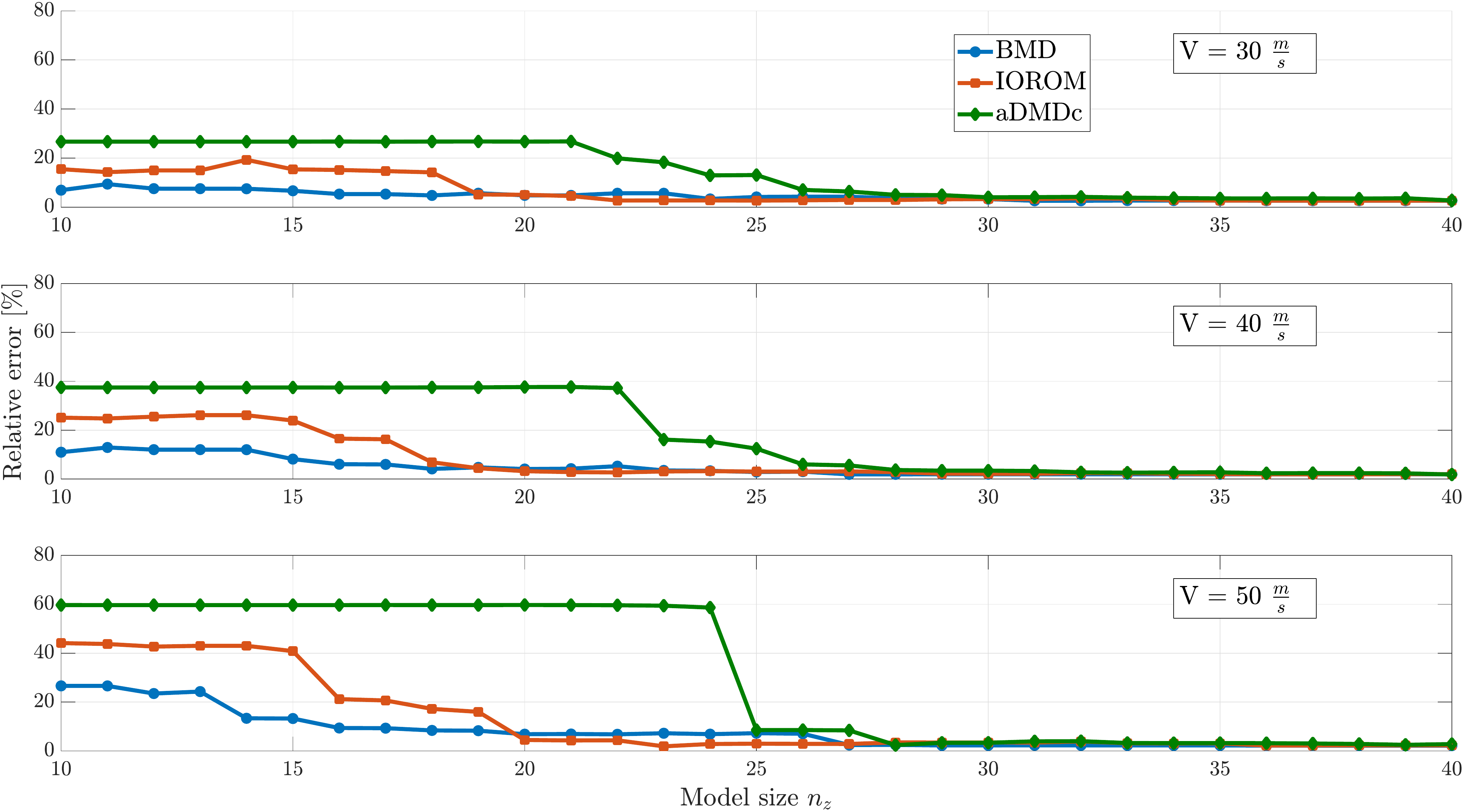}
       \caption{Relative error on the prediction of the system's output (bending mode) for three different values of the flight speed as a function of the model order.}
        \label{NewPlot_ErrvsROM}
\end{figure}

In this first set of tests, the accuracy of the different models at fixed values of the flight speed $V$ is assessed by means of sinusoidal inputs.
\begin{color}{black}
The frequencies, different for each channel, are expressed in terms of the aircraft reduced frequency $f_r=:\frac{V}{\bar{c}}$, where $\bar{c}$=0.29 m  is the mean chord of the wing. The first, third, and fifth input channels are excited with sinusoids $\frac{1}{5 \pi}f_r$, $\frac{1}{10 \pi}f_r$, and $\frac{1}{20 \pi}f_r$, respectively, while the other channels are set to zero. For reference, the first bending mode of the wing has a frequency of approximately 10 Hz \cite{Fasel_AIAA2019}.
\end{color}
This test is performed for 3 flight speeds in the range of operating conditions of interest, namely $V$=30 m/s, $V$=40 m/s, and $V$=50 m/s. To quantify the accuracy as a function of the order of the model $n_z$, the Euclidean norm of the error signal between the first bending mode amplitude provided by the high-fidelity FSI and each of the predictions obtained with the three ROM algorithm is computed. The results in terms of relative error (with respect to the predicated signal norm) are shown in Fig. \ref{NewPlot_ErrvsROM}.

It is clear that for all speeds BMD provides the smallest error for a very low-order approximation of the full dynamics, as expected in view of the choice of low-dimensional subspace where the high-dimensional data are projected. As the size $n_z$ of the system increases, the difference between the algorithms is less noticeable and, for high enough orders, the algorithms tend to give same results.

It is also noted that the results obtained with the aDMDc algorithm showed great sensitivity, in the range of $n_z$ displayed in Fig. \ref{NewPlot_ErrvsROM}, to the SVD truncation order $r$ employed in Eq. (\ref{Q_DMDc}). Using the hard threshold criterion from \cite{Donoho_thresh_2014} did not provide good results as it resulted in a very large $r$  (therefore the truncation included very low singular values deteriorating the approximation).
Since fine-tuning the value of $r$ to optimize the results obtained with aDMCc would have required trying several values of $r$ for each different value of $n_z$, this was not pursued here. Instead, the heuristic choice $r$=$n_z$+10 was implemented and proved to provide reasonable results. Even though, because of this possibly suboptimal choice, the resulting gap in performance with the other two methods observed in Fig. \ref{NewPlot_ErrvsROM} can be also ascribed to numerical inaccuracies associated with the decomposition (\ref{Q_DMDc}), the need to optimally choose $r$ can be considered as a disadvantage of the aDMDc problem formulation. It is finally observed that, in the extensive analyses performed, aDMCc typically showed worse performance than BMD irrespective of the choice of $r$.

\subsection{Parameter-varying models}\label{Results_varV}
In the second set of tests, the accuracy during parameter-varying manoeuvres is tested. A manoeuvre of 3 s where the flight speed linearly increases from  $V$=20 m/s to $V$=50 m/s is analyzed. Unless otherwise specified, the reduced-order models are obtained using snapshot matrices obtained gridding the flight speed range every 2 m/s and thus using 16 different speeds ($n_g$=16). A linear interpolation will be used to evaluate quantities for values of $\rho$ that are not in the grid.

\subsubsection{Sinusoidal excitation}\label{Results_varV_sine}
\begin{figure}[!h]
    \centering
        \includegraphics[width=1\columnwidth]{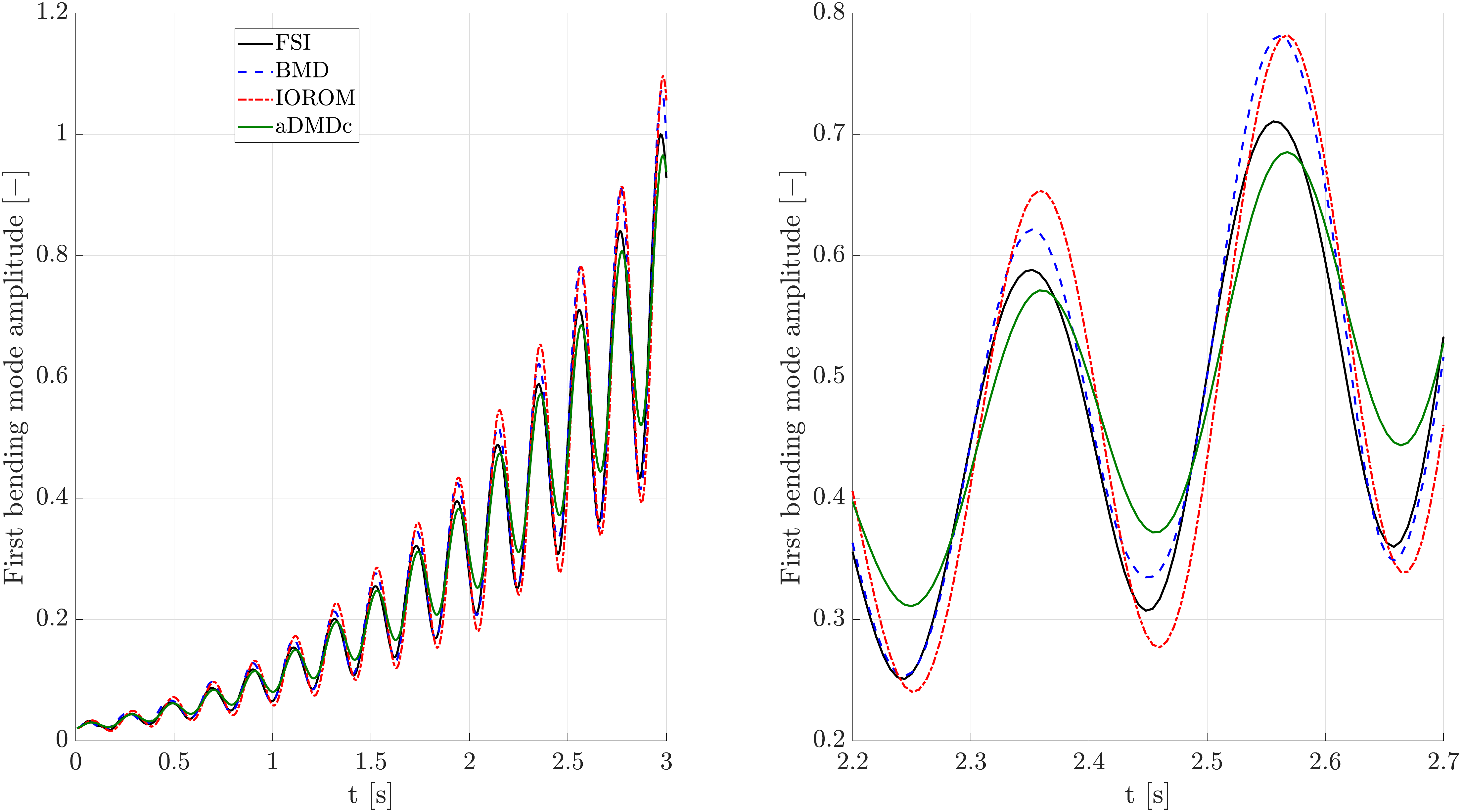}
       \caption{Comparison of the normalized output (bending mode) for parameter-varying simulations (flight speed linearly increasing from 20 m/s to 50 m/s) of models with size 14.}
        \label{LPV_bending}
\end{figure}

The same sinusoidal input signals used in Section \ref{Results_fixedV} are considered here. 
In Fig. \ref{LPV_bending}, the bending mode amplitude response obtained with the FSI solver (\emph{FSI}) is compared with the predictions of the three algorithms when the order of the models is fixed at $n_z$=14. \begin{color}{black}The goal is to compare the prediction accuracy of the algorithms when a very low number of states (in comparison with the order of the original system) is employed. It is noted that the same observations can be gathered when values of $n_z$ in the same range are considered.\end{color}
\begin{color}{black} All the signals are normalized by the largest value of the bending amplitude measured in the FSI simulation, which for the analyzed case was 0.7.  Note that a unitary value of the first bending mode corresponds to a wingtip displacement of 4.6 cm.\end{color}
In this case, since only one value of $n_z$ was considered, the aDMDc model was here obtained by fine tuning the threshold value $r$ in order to provide the best results. 

The plot confirms, also in the LPV setting, that the BMD algorithm guarantees the smallest error when a low-rank approximation of the system is desired.
In this simulation, aDMDc outperforms IOROM, possibly due to the fact that it uses a parameter-varying set of POD modes. However, aDMDc does not provide a family of interpolated low-order models, and interpolates directly the high-order states, thus requiring parallel simulations of the low-order models. The better performance of BMD, despite the fact that a part of the projection (the one related to the basis space) is constant, is ascribed to the improved selection of subspace for the projection compared to the standard POD one, common to the other two methods. In addition to the improvement in the accuracy, the BMD algorithm is also capable of providing, like the IOROM algorithm, a family of consistent LTI models with the advantages for LPV control design and in general real-time applications.


\subsubsection{Effect of the input signals}
\begin{figure}[!h]
    \centering
        \includegraphics[width=1\columnwidth]{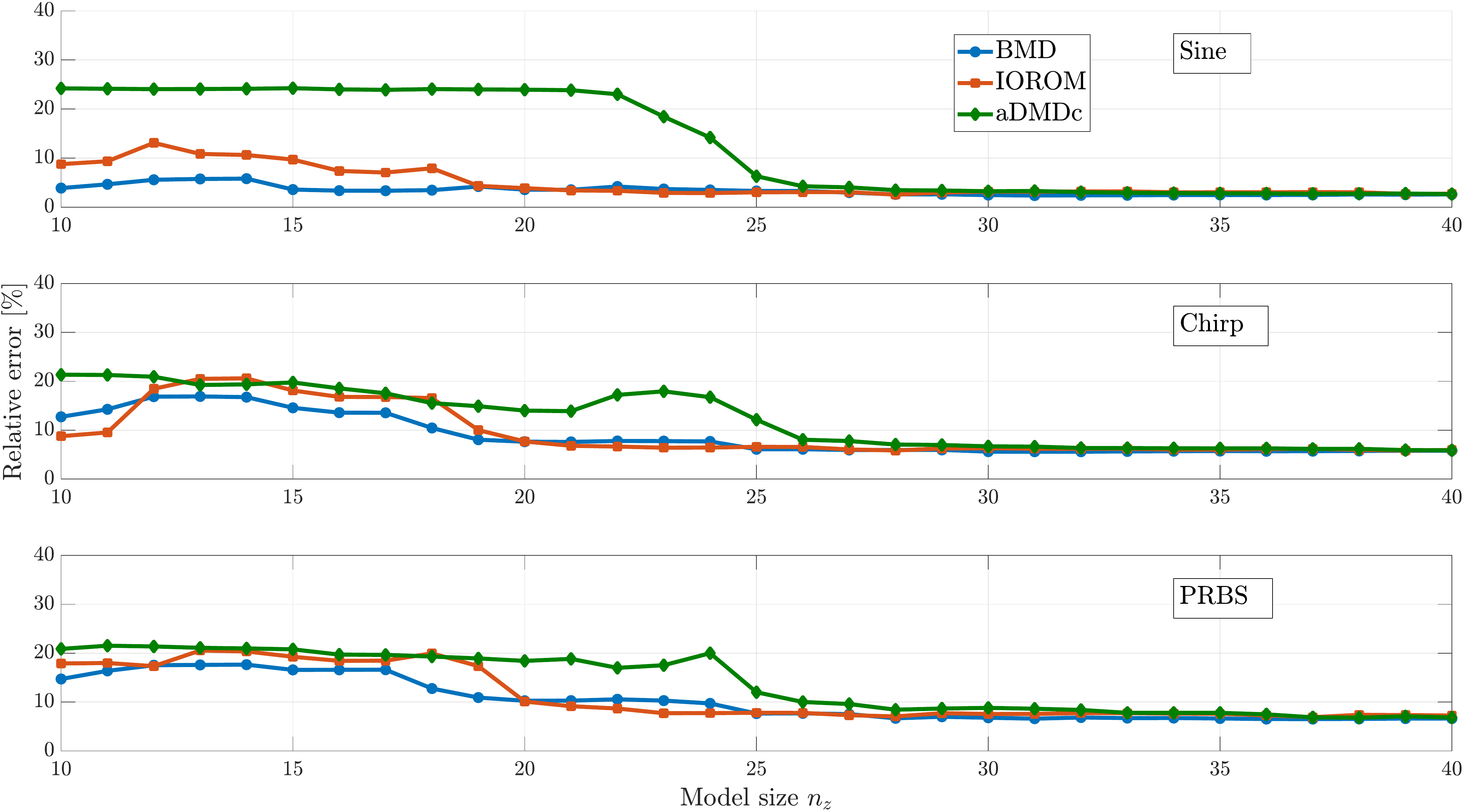}
       \caption{Relative error of the prediction of the system's output (bending mode) for speed-varying manoeuvres with three different types of input signals as a function of the model order.}
        \label{ErrvsROM_LPV_input}
\end{figure}

This section investigates the accuracy of the reduced-order models for different types of input signals. The Euclidean norm of the error signal between the first bending mode amplitude provided by the high-fidelity FSI and the prediction obtained with each of the three ROM algorithms is again used as metric to assess the quality of the approximation. Three classes of inputs are considered: \emph{Sine} coincides with the signal tested so far and already investigated in \cite{Fonzi_Royal2020}; \emph{Chirp} excites the system by injecting in all 6 input channels defined in (\ref{input_channels}) a chirp signal with frequency linearly varying from $\frac{1}{50 \pi}f_r$ to $\frac{2}{15 \pi}f_r$; \emph{PRBS} excites the system by injecting in all 6 input channels a PRBS-9 sequence. This last input, namely a Pseudo-Random Binary Signal (PRBS) \cite{Ljung1999}, is a deterministic signal with white-noise-like properties. It is very well known in the system identification and experiment design fields since it has the favourable property of equally distributing energy across all the frequencies in the input spectrum. In this way, information on the models in different frequency ranges can be extracted. Although not a common input in aeroservoelastic applications, it has been used in this spirit here, since the previously adopted sets of input will only give information on the behaviour of the reduced models around the aircraft reduced frequency $f_r$.
Results are shown in Fig. \ref{ErrvsROM_LPV_input}.

The plots confirm the advantage in using the BMD approach when seeking a low-order model capturing parameter variations. These results are important considering that they are obtained by exploring different frequency ranges of the system's response.

\begin{color}{black}
An interesting aspect observed in Fig. \ref{ErrvsROM_LPV_input} is that none of the algorithms exhibit a monotonic improvement of the model's accuracy (measured here by the relative prediction's error) as a function of the system’s order. Indeed, in a very few cases, a small deterioration can be observed between two consecutive values of $n_z$, before the curve keeps decreasing as $n_z$ increases. Even though this might seem surprising at first glance, there is no theoretical guarantee that such a monotonic improvement is achieved in this type of algorithms. The reason is that adding one mode to the low dimensional subspace where the dynamics is described might (in principle) deteriorate the model's approximation, if that mode alone does not add meaningful information. This is for example the case when pairs of modes describe relevant features of the system (e.g. modes of vibration), and so only when both of them are used in the projection there is an improvement. In practice, numerical errors due to the SVD truncation and the interpolation of the models across the parameter's grid can also have an effect on these results. However, it is noted that the number of occurrences and the entity of this deterioration are limited.
\end{color}

\subsubsection{Effect of the parameter grid}
\begin{figure}[!h]
    \centering
        \includegraphics[width=1\columnwidth]{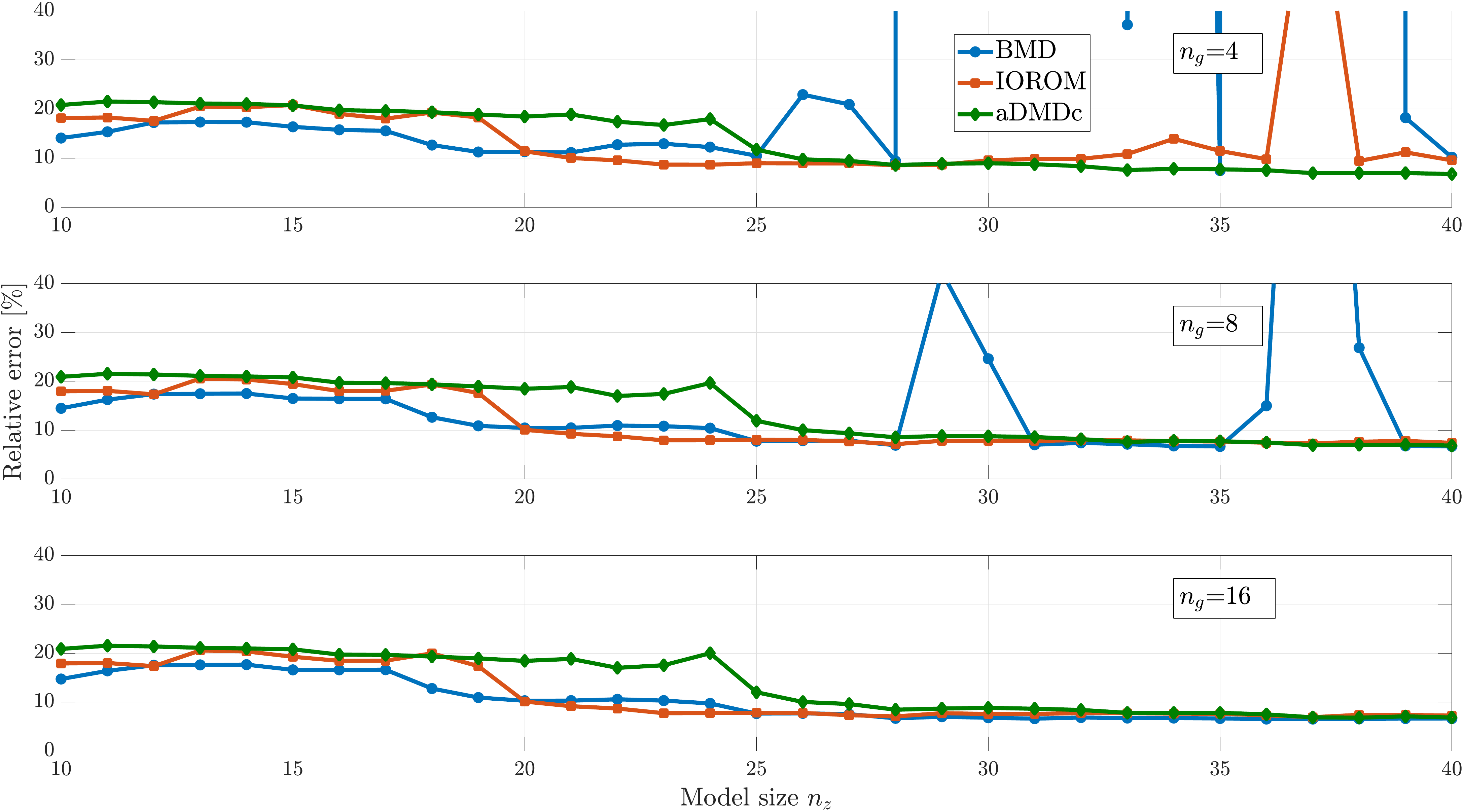}
       \caption{Relative error on the prediction of the system's output (bending mode) for speed-varying manoeuvres with values of $n_g$ as a function of the model order.}
        \label{ErrvsROM_LPV_ng}
\end{figure}

In this section we analyze the effect of the flight speed grid where the reduced-order LTI models are computed, i.e. the selection of the parameter $n_g$. This is an important aspect, which is known to influence both the accuracy of the LPV models and the quality of the control design based on them. Three cases are compared in Fig. \ref{ErrvsROM_LPV_ng}: $n_g$=4 where the grid includes one plant every 10 m/s; $n_g$=8 where the grid includes one plant every 4 m/s from $V$=20 m/s to $V$=48 m/s and then $V$=50 m/s; $n_g$=16 which is the grid used so far. PRBS inputs are used to excite the system in these tests.

From the analyses it can be gathered that aDMDc is more robust than the other algorithms to the value of $n_g$. In particular, both IOROM and BMD present poor performance for a few reduced order models in the range of $n_z$ between 30 and 40 when the flight speed grid is coarser. The reason for this behaviour is due to the interpolation approaches employed by the three ROM schemes. Whereas IOROM and BMD interpolate the low-order state-space matrices obtained at the grid points, aDMDc interpolates directly the high-order vector states which are obtained by lifting the low-order states $\tilde{z}$ from the local models (\ref{aDMDc_ROM_f}) running in parallel.
Interpolating every entry of the state-space matrices therefore makes the choice of the grid a more delicate aspect in IOROM and BMD. The unstable behaviours resulting in very high (out of the plot) errors are indeed ascribed to numerical inaccuracies in this interpolation. It has been observed that the entries of the matrices are overall bigger as the order $n_z$ is increased, hence justifying why these outliers take place within the aforementioned range of model's orders.
\begin{color}{black}
While there does not seem to be a fundamental reason to explain it, it is apparent that the sensitivity of BMD to the coarseness of the parameter grid is more accentuated. A possible explanation is that, because at each grid point the computation of the empirical Gramians is required, dealing with coarse grid exacerbates the numerical inaccuracies associated with the projections on low-order models. Improved interpolation schemes, not considered in this work, could be employed to ameliorate this issue.
\end{color}
Except for these isolated numerical problems, BMD shows better performance even when very coarse grids are employed.

\subsubsection{Prediction capability for other signals}
The capability of the models to predict other quantities of interest, such as for example aerodynamic coefficients depending on the system's states, is investigated.
\begin{color}{black}
In particular, we test the accuracy when these coefficients are added to the vector of output (this is done for BMD) or computed directly from the states (this is done for IOROM and aDMDc).
In the latter case, the low-order states are lifted to the high-order ones, which are then used to compute the coefficients using their known relationship to the states. While this is the only possible way of reconstructing the system's signals for aDMDc, in IOROM and BMD this can alternatively be done by simply adding the desired quantities to the vector of outputs. This would probably be the preferred approach if the signals are used for control (either because they represent measurements fed to the controller or because they are performance measures to be optimized). The different choice done here for BMD and IOROM is for the sake of exploring different models, and results showed that whether the signals were computed from output channels or retrieved from the states had a very minor impact on the predictive accuracy.
\end{color}

\begin{figure}[!h]
    \centering
        \includegraphics[width=1\columnwidth]{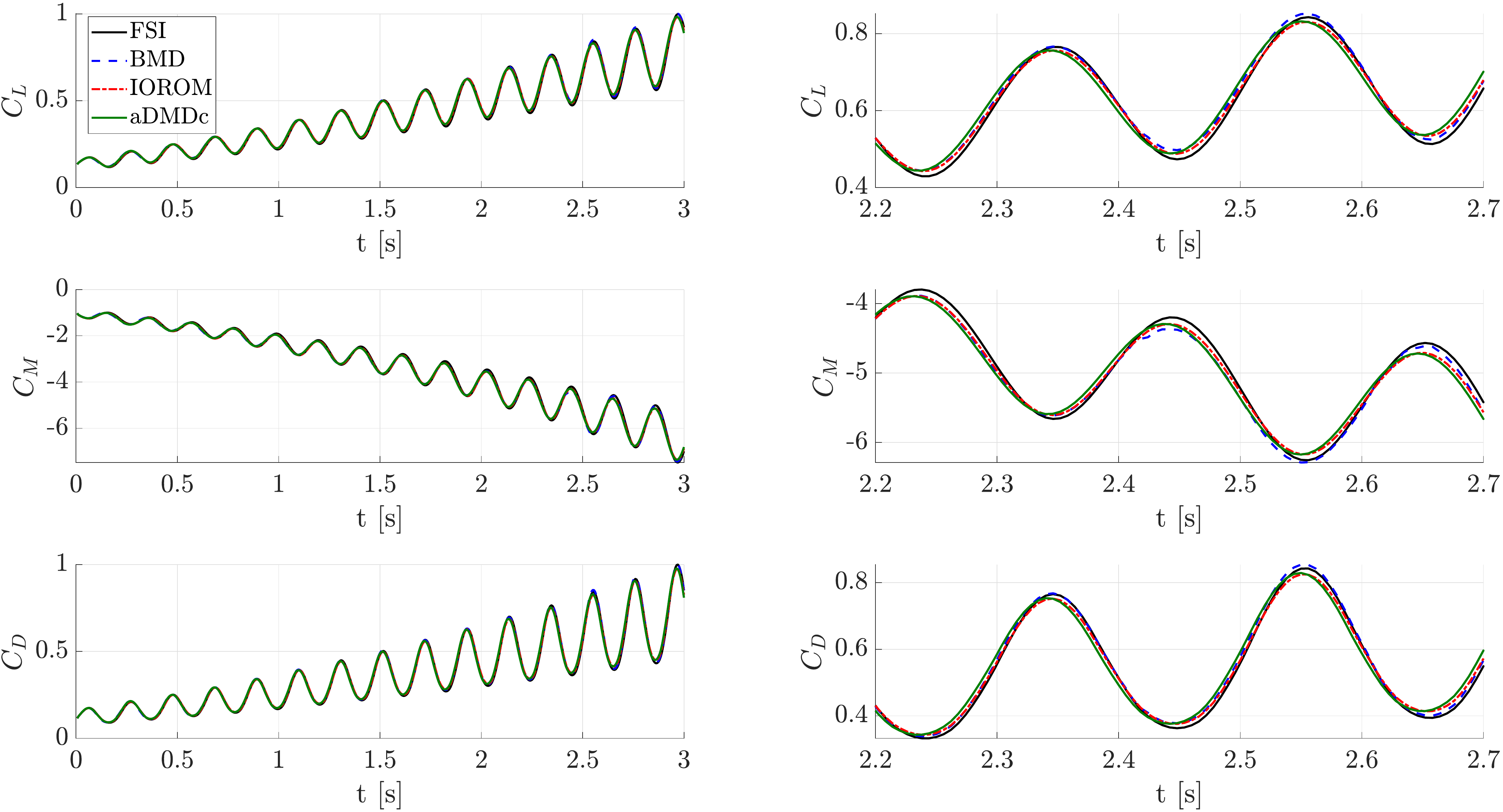}
       \caption{Comparison of the normalized lift, pitch, and drag coefficients for a parameter-varying simulation with $n_z$=14 and sinusoidal inputs.}
        \label{LPV_ClPitch_cD}
\end{figure}

Figure \ref{LPV_ClPitch_cD} shows the normalized lift ($C_L$), pitch ($C_M$), and drag  ($C_D$) coefficients for the same constant acceleration manoeuvre considered in the previous sections and with a sinusoidal excitation. Normalization is performed, as done earlier in Figure \ref{LPV_bending}, by dividing each signal by the largest value of the corresponding signal in the FSI simulation.

\begin{figure}[!h]
    \centering
        \includegraphics[width=1\columnwidth]{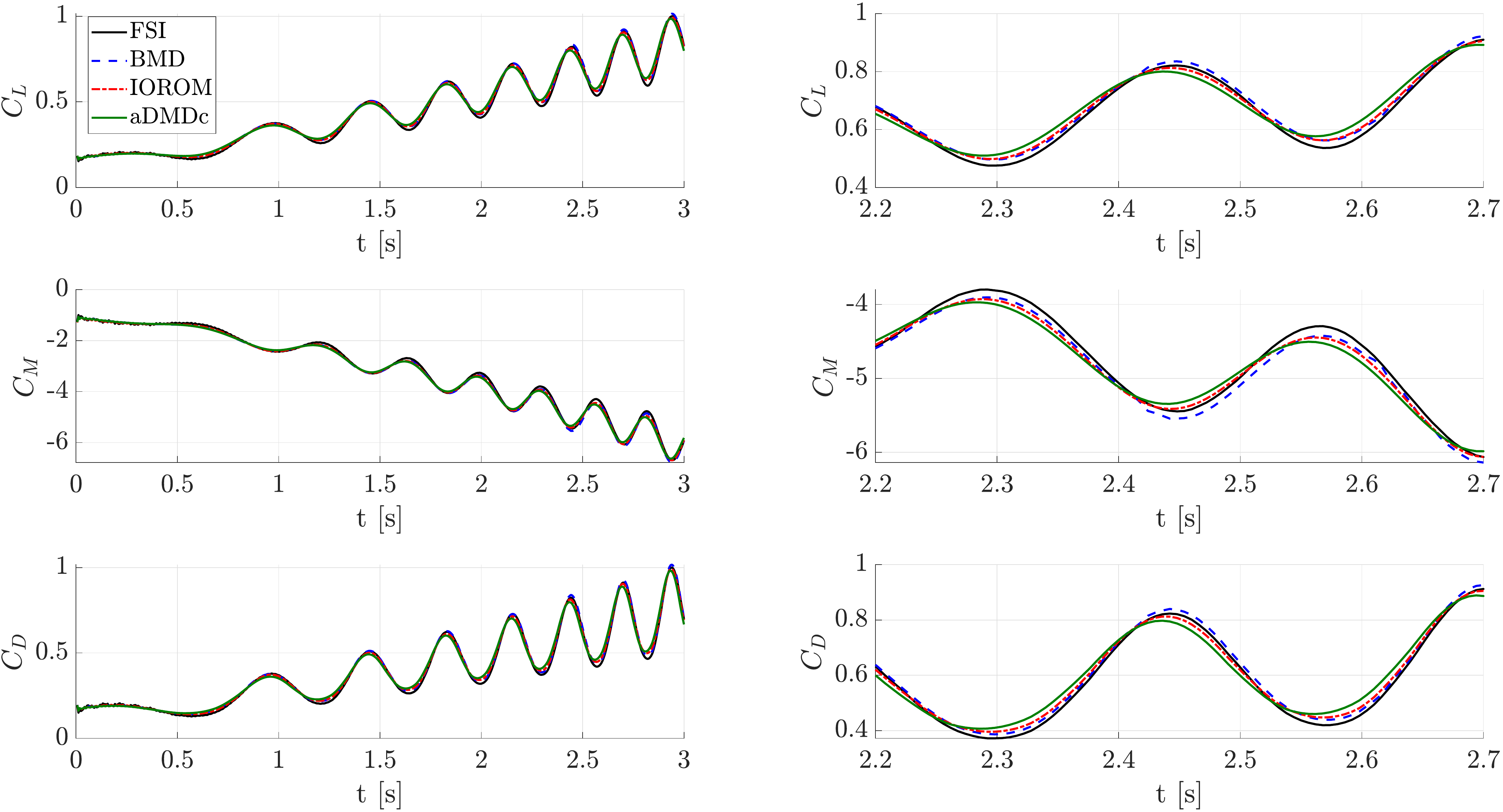}
       \caption{Comparison of the normalized lift, pitch, and drag coefficients for a parameter-varying simulation with $n_z$=14 and chirp inputs.}
        \label{LPV_ClPitch_cD_chirp}
\end{figure}

The same observations gathered earlier with respect to the trajectory of the bending mode are confirmed here. It is particularly interesting to observe that, even though these coefficients are not outputs of the model, and thus the balancing projection is not aimed directly at capturing them, the BMD algorithm is still able to perform better than the others. Figure \ref{LPV_ClPitch_cD_chirp} shows the same analyses when chirp signals are used as input to excite the model. Similar conclusions can be drawn.

\subsection{Reduced-order models for Model Predictive Control}\label{Results_MPC}
In this section, a control application of the morphing wing's low-order models is investigated. Specifically, model predictive control (MPC) \cite{rawlings2009model} is considered, given its well established use in the AWE field \cite{Fagiano_IJRNC}.
Two distinct reference tracking problems are examined, where pre-defined lift and first bending mode amplitude profiles are tracked while flying trajectories over a range of different flight speeds and in the presence of turbulence and gusts. The analysis of these manoeuvres is motivated by the interest in using active control to guarantee a safe operation for the AWE system (with respect to some of its critical components such as the wing or tether) by keeping indicators of the structural integrity close to desired, and possibly pre-optimized, values. This can avoid passive remedies such as reducing the load transmitted to the ground station, which in turn decreases the amount of wind energy harvested. Having effective and reliable control laws to guarantee the integrity of the AWE system can represent an important enabler for this technology \cite{Schmehl_bookAWE2018}.

In its basic form, model predictive control repeatedly solves a finite-horizon optimal control problem of length $N_c$ subject to input and state-constraints. At each instant, a model of the system is employed to predict its response and thus select the control sequence $(u_i)_{i=0}^{N_c-1}$ which minimizes the cost
\begin{equation}\label{MPC_cost}
    J_\text{MPC}=\sum_{k=0}^{N_c-1}\left(\norm{\tilde{y}_k-r_{k}}_N^2+\norm{\tilde{u}_k}_M^2+\norm{\Delta \tilde{u}_k}_{M_{\Delta}}^2\right),
\end{equation}
where $r$ is the reference trajectory, and for a vector $x$, we denote by $\norm{x}_P$ the weighted $l_2$-norm $(x^\mathsf{T}Px)^{\frac{1}{2}}$.
Besides the terms penalizing deviation of the output $\tilde{y}$ from $r$ (with the weighting matrix $N \in \mathbb{R}^{n_y \times n_y}$) and control effort (with the weighting matrix $M \in \mathbb{R}^{n_u \times n_u}$), the cost in (\ref{MPC_cost}) also penalizes fast changes in the input via the term $\Delta \tilde{u}_k=\tilde{u}_k-\tilde{u}_{k-1}$ (e.g. to avoid actuator rate saturation).

The following optimization problem will be solved to obtain the optimal input sequence
\begin{subequations}\label{MPC_opt}
\begin{align}
    \underset{(u_i)_{i=0}^{N_c-1}, (y_i)_{i=0}^{N_c-1}}{\text{minimize}}  \quad & J_\text{MPC}(\tilde{u},\tilde{y}), \label{MPC_opt_1}\\
    \text{subject to} \quad & \tilde{y}_i=f(\tilde{u}_i,\tilde{u}_{i+1}), \label{MPC_opt_2}\\
    & (\tilde{u}_i)_{i=0}^{N-1} \in \mathcal{U}, \label{MPC_opt_3}
\end{align}
\end{subequations}
where: the cost function (\ref{MPC_opt_1}) is defined in (\ref{MPC_cost}); the constraint set $\mathcal{U}$ (\ref{MPC_opt_3}) enforces minimum and maximum values for the input; and (\ref{MPC_opt_2}) enforces the dynamic constraint that relates the sequence of input to the output via an input-output model of the system $f$. Precisely, $f$ will be formulated here by using the reduced-order aDMDc, IOROM, and BMD models.
The goal of the analyses is to compare the associated closed-loop cost $J_\text{MPC}^\text{CL}$, that is the cost (\ref{MPC_cost}) incurred by the true system (simulated here by the high-dimensional FSI solver) when this is regulated by the inputs optimized solving problem (\ref{MPC_opt}). Since the model is used to predict the system's output, and thus select the input sequence, any mismatch between model and system can result in a degradation of the controller performance.

\begin{figure}[!h]
    \centering
        \includegraphics[width=1\columnwidth]{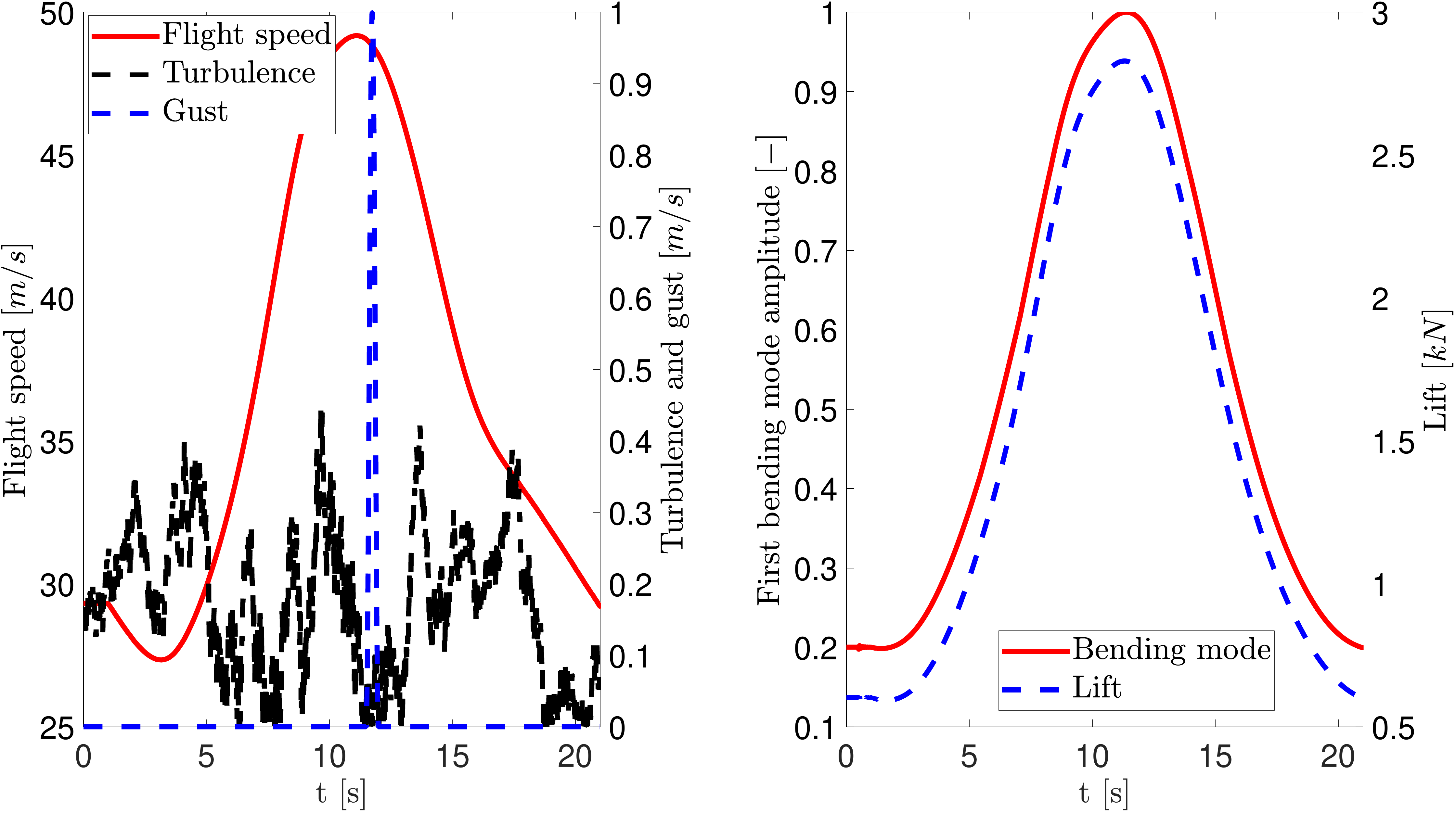}
       \caption{Speeds variations during the manoeuvres (left plot) and reference tracking profiles (right plot).}
        \label{MPC_data}
\end{figure}

The analysis considers the case where the morphing wing (Fig. \ref{Figure1}, I) flies a trajectory with flight velocities ranging between $V$=27 m/s and $V$=50 m/s (Fig. \ref{MPC_data}-left). Additionally, the wing is subject to a gust at the maximum flight speed with gust length 0.5 s corresponding to a 1 degree deflection of the angle of attack $\alpha$, and to turbulence generated with a Dryden filter (Fig. \ref{MPC_data}-left). The right plot in Fig. \ref{MPC_data} depicts the lift and the first bending mode's amplitude profiles tracked by the MPC algorithm. Recall from the previous discussion that a unitary value of the first bending mode corresponds to a wingtip displacement of 4.6 cm.

The scenario considered here has only a symmetric morphing actuation input, i.e. $\tilde{u}=F_s$. This normalized input is constrained to be in the interval \newline [-3, 3] at each time-step. This is associated with allowable deflections of the trailing edge in the range $\pm$ 9 mm. Problem (\ref{MPC_opt}) is solved using the MATLAB implementation provided in \cite{Kaiser_MPC18}, where the application of MPC with models obtained via DMDc was investigated.

The output $\tilde{y}$ is either the first bending mode, or the lift force generated by the wing, depending on the case. The control horizon is $N_c$=10 and the weights used in the cost are: $N$=1300, $M$=10, and $M_{\Delta}$=0.1 (lift tracking) and $N$=13000, $M$=10, and $M_{\Delta}$=0.1 (bending tracking). The penalty on the output deviation is increased in the latter case due to the difference in magnitude of the two tracked quantities (recall Fig. \ref{MPC_data}-right).
\begin{color}{black}
Figure \ref{MPC_track1} shows the comparison of the closed-loop cost $J_\text{MPC}^\text{CL}$ resulting from closing the true plant (simulated by means of the high-dimensional FSI solver) with the MPC controller generated using the low-order models.
\end{color}
For the sake of clarity, the closed-loop cost $J_\text{MPC}^\text{CL}$ has been normalized in each case by dividing it by the corresponding value obtained with the BMD algorithm when $n_z$=40.
\begin{figure}[h!]
\begin{center}
\subfigure[Lift.]{
\includegraphics[width=0.75\columnwidth]{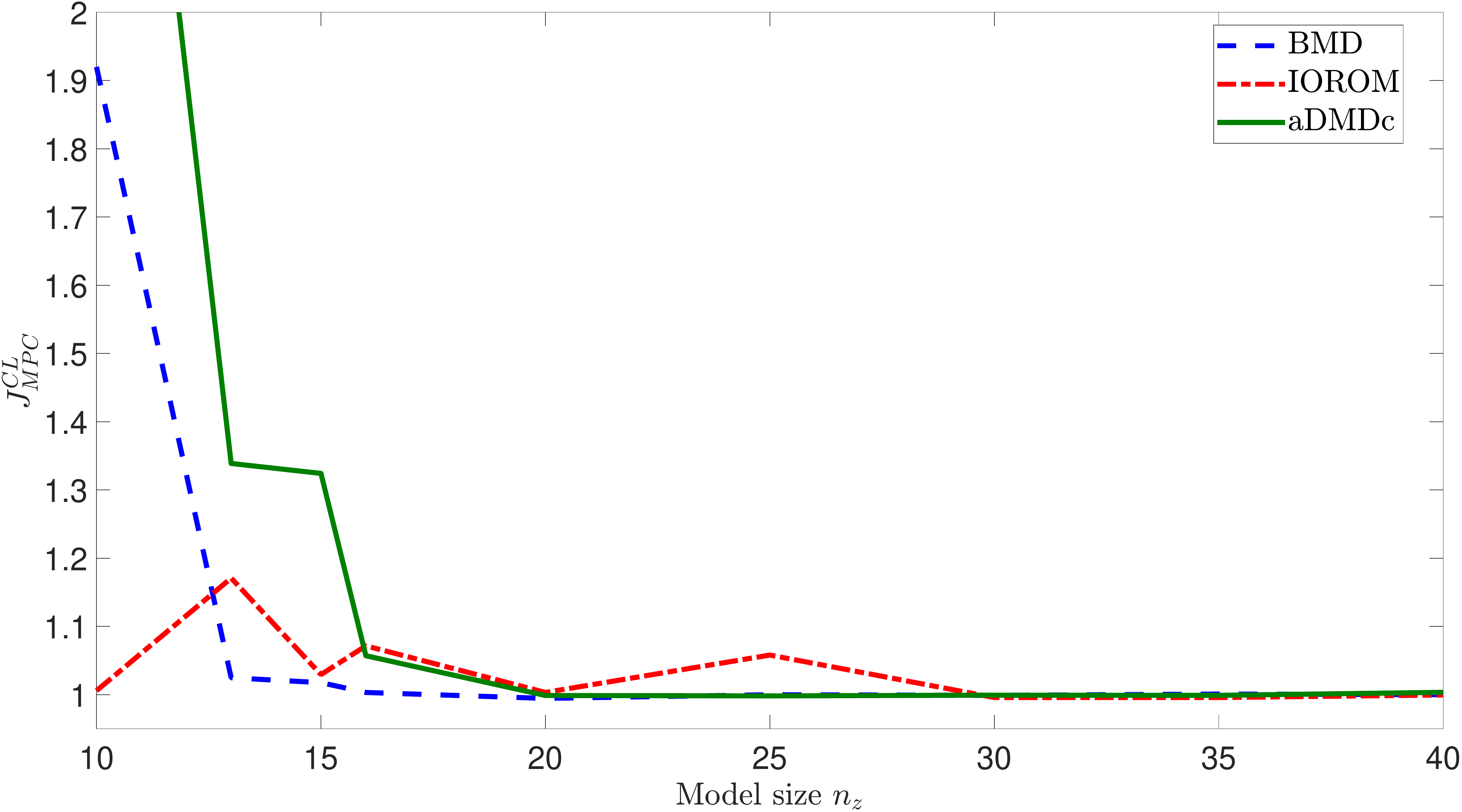}
\label{MPC_track1}}
\subfigure[First bending mode.]{
\includegraphics[width=0.75\columnwidth]{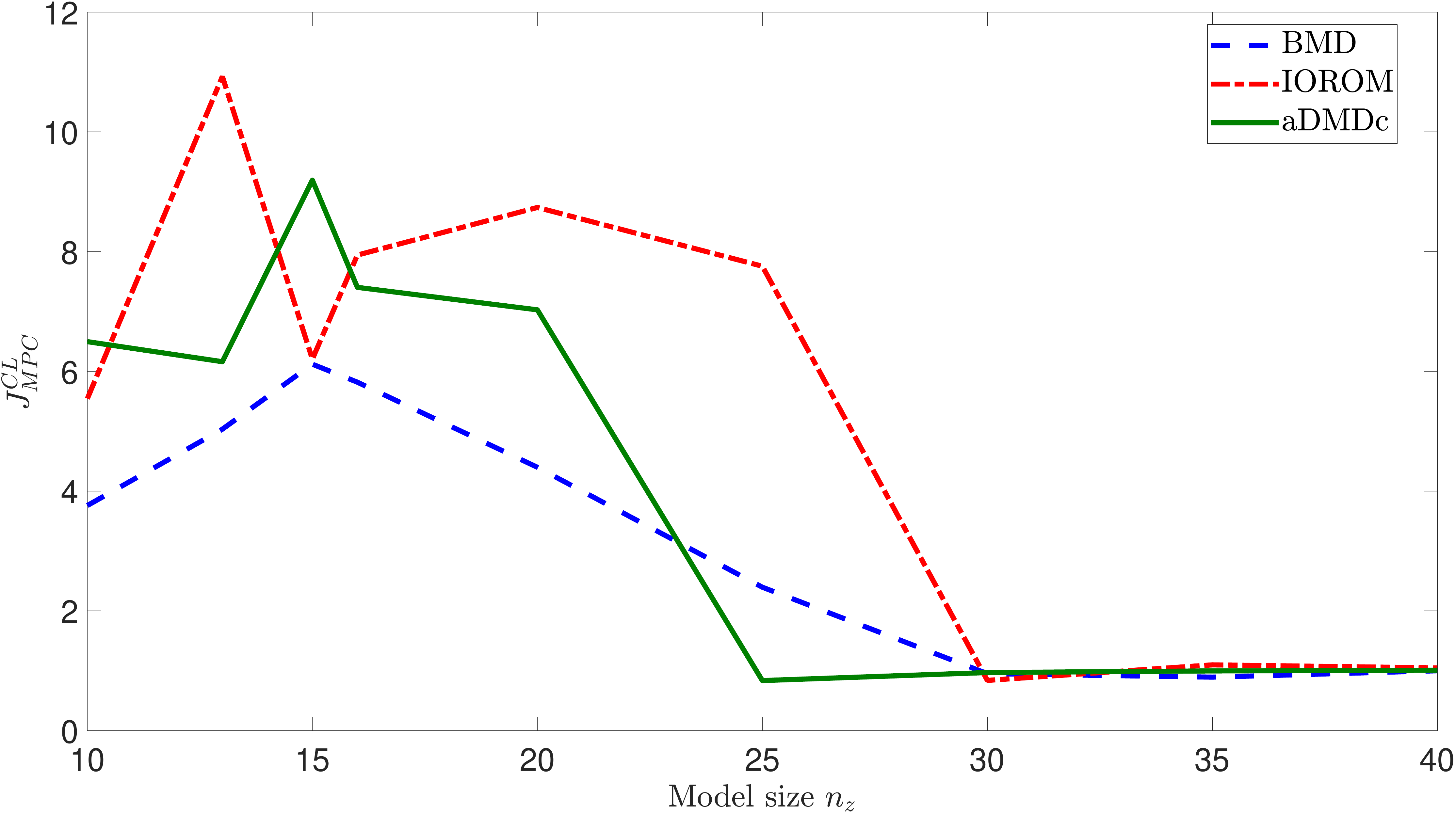}
\label{MPC_track2}}
\caption{Normalized closed loop costs for the two MPC tracking problems as a function of the model order.}
\label{MPC_track1}
\end{center}
\end{figure}

The observations gathered in the previous sections regarding the better prediction performance achieved with the BMD algorithm when low-order approximations are considered are confirmed here in the context of control applications. Both plots show that, while for higher orders the closed-loop costs have very similar values, when the size of the model is decreased the BMD gives in general the lowest cost. Another interesting observation is that the lift tracking problem is quite robust to the use of low-order models. Indeed, the closed-loop costs are always within two times of the lowest cost achieved at $n_z$=40 except for the case of aDMDc at $n_z$=10). On the other hand, the bending tracking problem is shown to be more challenging when low-order representations are employed. Whereas no attempt to further optimize the MPC problem tuning was made (all the design parameters were kept the same independent of $n_z$), this motivates further work on the use of low-order models for control of coupled flexible structures like those encountered in AWE applications. In the real-time control setting, it is important to stress that by using BMD and IOROM models an order of magnitude computational speed-up was achieved with respect to the cases where aDMDc models were used. This is because the aDMDc models have the requirement of running several models in parallel.

%
%
%

\section{Conclusion}
The paper proposes the Balanced Mode Decomposition with oblique projection algorithm, a novel data-driven algorithm for constructing low-order LPV models from system's trajectories. Two recent algorithms from the literature, aDMDc and IOROM, are considered for comparison since they both have connections with the newly proposed approach. Technical details on the BMD algorithm are given in order to clearly point out the innovations, and the advantages with respect to previous work. The performance of the BMD algorithm is assessed on a morphing wing for airborne wind energy applications. The results, proposed both for the fixed parameter and, more extensively, for the parameter-varying case, confirm the theoretical advantages discussed in the technical part of the paper. When seeking low-order model representations, the BMD approach achieves generally, among the tested algorithms, the lowest prediction error and best control performance when used as model for an on-line MPC scheme. The improved accuracy is ascribed to the use of a projecting subspace that balances the low-order states (this element is of interest also in a fixed-parameter setting), and to the use of a parameter-varying projection operator (which can thus be enriched with parameter-dependent features, instead of being fixed throughout the range). This has the advantageous feature of being achieved while guaranteeing state-consistency. \begin{color}{black}Owing to these appealing features, it is envisaged the application of BMD for tasks such as off-line and real-time control design, and in multi-disciplinary optimization tool chains, where typically low-order representations are employed as surrogate models.\end{color}



\section*{Acknowledgements}
This work is supported by the Swiss National Science Foundation under grant no. $200021\_178890$. The authors wish to thank Nivethan Yogarajah and Eva Ahbe for helpful discussions.



\section*{Declaration of competing interest}
The authors declare that they have no conflict of interest.

\bibliography{BalancedLPV_AST}

\end{document}